\documentclass[10pt,twocolumn]{article}
\usepackage{graphicx}
\usepackage{amsmath,amssymb,cite,times}

\DeclareMathOperator{\diag}{diag}
\newcommand{\bg}[1]{\boldsymbol{#1}} 
\newcommand{\bm}[1]{\mathbf{#1}} 

\newcommand\T{{\mathpalette\raiseT\intercal}}
\newcommand\raiseT[2]{%
\setbox0\hbox{$#1{#2}$}\raise\dp0\box0}

\usepackage{algorithm,algorithmic}
\makeatletter
\newcommand{\ALOOP}[1]{\ALC@it\algorithmicloop\ #1%
  \begin{ALC@loop}}
\newcommand{\ENDALOOP}{\end{ALC@loop}\ALC@it\algorithmicendloop}

\makeatother

\title{\Large\textbf{Shape Classification using Spectral Graph Wavelets}}
\author{Majid Masoumi and A. Ben Hamza\\
Concordia Institute for Information Systems Engineering\\
Concordia University, Montreal, QC, Canada 
}
\date{}

\begin{document}
\maketitle

\begin{abstract}
Spectral shape descriptors have been used extensively in a broad spectrum of geometry processing applications ranging from shape retrieval and segmentation to classification. In this paper, we propose a spectral graph wavelet approach for 3D shape classification using the bag-of-features paradigm. In an effort to capture both the local and global geometry of a 3D shape, we present a three-step feature description framework. First, local descriptors are extracted via the spectral graph wavelet transform having the Mexican hat wavelet as a generating kernel. Second, mid-level features are obtained by embedding local descriptors into the visual vocabulary space using the soft-assignment coding step of the bag-of-features model. Third, a global descriptor is constructed by aggregating mid-level features weighted by a geodesic exponential kernel, resulting in a matrix representation that describes the frequency of appearance of nearby codewords in the vocabulary. Experimental results on two standard 3D shape benchmarks demonstrate the effectiveness of the proposed classification approach in comparison with state-of-the-art methods.
\end{abstract}

\section{Introduction}
The recent surge of interest in the spectral analysis of the Laplace-Beltrami operator (LBO)~\cite{Rosenberg:97} has resulted in a glut of spectral shape signatures that have been successfully applied to a broad range of areas, including object recognition and deformable shape analysis~\cite{Levy:06,Reuter:06,Rustamov:07,Bronstein:11,Chunyuan:13b,Biasotti:SHREC17,Rodola:SHREC17}, multimedia protection~\cite{Tarmissi:09}, and shape classification~\cite{Gao:14}. The diversified nature of these applications is a powerful testimony of the practical usage of spectral shapes signatures, which are usually defined as feature vectors representing local and/or global characteristics of a shape and may be broadly classified into two main categories: local and global descriptors. Local descriptors (also called point signatures) are defined on each point of the shape and often represent the local structure of the shape around that point, while global descriptors are usually defined on the entire shape.

Most point signatures may easily be aggregated to form global descriptors by integrating over the entire shape. Rustamov~\cite{Rustamov:07} proposed a local feature descriptor referred to as the global point signature (GPS), which is a vector whose components are scaled eigenfunctions of the LBO evaluated at each surface point. The GPS signature is invariant under isometric deformations of the shape, but it suffers from the problem of eigenfunctions' switching whenever the associated eigenvalues are close to each other. This problem was lately well handled by the heat kernel signature (HKS)~\cite{Sun:09}, which is a temporal descriptor defined as an exponentially-weighted combination of the LBO eigenfunctions. HKS is a local shape descriptor that has a number of desirable properties, including robustness to small perturbations of the shape, efficiency and invariance to isometric transformations. The idea of HKS was also independently proposed by G\c ebal~\textit{et al.}~\cite{Gebal:09} for 3D shape skeletonization and segmentation under the name of auto diffusion function. To give rise to substantially more accurate matching than HKS, the wave kernel signature (WKS)~\cite{Aubry:11} was proposed as an alternative in an effort to allow access to high-frequency information. Using the Fourier transform's magnitude, Bronstein and Kokkinos~\cite{Kokkinos:10} introduced the scale invariant heat kernel signature (SIHKS), which is constructed based on a logarithmically sampled scale-space.

One of the simplest spectral shape signatures is Shape-DNA~\cite{Reuter:06}, which is an isometry-invariant global descriptor defined as a truncated sequence of the LBO eigenvalues arranged in increasing order of magnitude. Gao~\textit{et al.}~\cite{Gao:14} developed a variant of Shape-DNA, referred to as compact Shape-DNA (cShape-DNA), which is an isometry-invariant signature resulting from applying the discrete Fourier transform to the area-normalized eigenvalues of the LBO. Chaudhari~\textit{et al.}~\cite{Chaudhari:14} presented a slightly modified version of the GPS signature by setting the LBO eigenfunctions to unity. This signature, called GPS embedding, is defined as a truncated sequence of inverse square roots of the area-normalized eigenvalues of the LBO. A comprehensive list of spectral descriptors can be found in~\cite{Lian:13,Chunyuan:14b}.

From the graph Fourier perspective, it can be seen that HKS is highly dominated by information from low frequencies, which correspond to macroscopic properties of a shape. Wavelet analysis has some major advantages over Fourier transform, which makes it an interesting alternative for many applications. In particular, unlike the Fourier transform, wavelet analysis is able to perform local analysis and also makes it possible to perform a multiresolution analysis. Classical wavelets are constructed by translating and scaling a mother wavelet, which is used to generate a set of functions through the scaling and translation operations. The wavelet transform coefficients are then obtained by taking the inner product of the input function with the translated and scaled waveforms. The application of wavelets to graphs (or triangle meshes) is, however, problematic and not straightforward due in part to the fact that it is unclear how to apply the scaling operation on a signal (or function) defined on the mesh vertices. To tackle this problem, Coifman and Lafon~\cite{Coifman:06} introduced the diffusion wavelets, which generalize the classical wavelets by allowing for multiscale analysis on graphs. The construction of diffusion wavelets interacts with the underlying graph through repeated applications of a diffusion operator, which induces a scaling process. Hammond~\emph{et al.}~\cite{Hammond:11} showed that the wavelet transform can be performed in the graph Fourier domain, and proposed a spectral graph wavelet transform that is defined in terms of the eigensystem of the graph Laplacian matrix. More recently, a spectral graph wavelet signature (SGWS) was introduced in~\cite{Chunyuan:13b,Chunyuan:13c,Masoumi:16}. SGWS is a multiresolution local descriptor that is not only isometric invariant, but also compact, easy to compute and combines the advantages of both band-pass and low-pass filters.

A popular approach for transforming local descriptors into global representations that can be used for 3D shape recognition and classification is the bag-of-features (BoF) model~\cite{Bronstein:11}. The task in the shape classification problem is to assign a shape to a class chosen from a predefined set of classes. The BoF model represents each shape in the training dataset as a collection of unordered feature descriptors extracted from local areas of the shape, just as words are local features of a document. A baseline BoF approach quantizes each local descriptor to its nearest cluster center using K-means clustering and then encodes each shape as a histogram over cluster centers by counting the number of assignments per cluster. These cluster centers form a visual vocabulary or codebook whose elements are often referred to as visual words or codewords. Although the BoF paradigm has been shown to provide significant levels of performance, it does not, however, take into consideration the spatial relations between features, which may have an adverse effect not only on its descriptive ability but also on its discriminative power. To account for the spatial relations between features, Bronstein~\textit{et al.} introduced a generalization of a bag of features, called spatially sensitive bags of features (SS-BoF)~\cite{Bronstein:11}. The SS-BoF is a global descriptor defined in terms of mid-level features and the heat kernel, and can be represented by a square matrix whose elements represent the frequency of appearance of nearby codewords in the vocabulary. In the same spirit, Bu~\textit{et al.}~\cite{Bu:14} recently proposed the geodesic-aware bags of features (GA-BoF) for 3D shape classification by replacing the heat kernel in SS-BoF with a geodesic exponential kernel.

In this paper, we propose a 3D shape classification approach, called SGWC-BoF, which employs spectral graph wavelet codes (SGWC) obtained from spectral graph wavelet signatures (i.e. local descriptors) via the soft-assignment coding step of the BoF model in conjunction with a geodesic exponential kernel for capturing the spatial relations between features. Shape classification ~\cite{Masoumi:17} is the process of organizing a dataset of shapes into a known number of classes, and the task is to assign new shapes to one of these classes. In addition to taking into consideration the spatial relations between features via a geodesic exponential kernel, the proposed approach performs classification on spectral graph wavelet codes, thereby seamlessly capturing the similarity between these mid-level features. We not only show that our formulation allows us to take into account the spatial layout of features, but we also demonstrate that the proposed framework yields better classification accuracy results compared to state-of-the-art methods, while remaining computationally attractive. The main contributions of this paper may be summarized as follows:
\begin{enumerate}
  \item We present local shape descriptors using multiresolution analysis of spectral graph wavelets.
  \item We construct mid-level features by embedding the local shape descriptors into the visual vocabulary space using the soft assignment coding step of the bag-of-features paradigm.
  \item We introduce a global descriptor, which is constructed by aggregating mid-level features weighted by a geodesic exponential kernel.
\end{enumerate}

The remainder of this paper is organized as follows. In Section 2, we briefly overview the Laplace-Beltrami operator and spectral signatures. In Section 3, we introduce a three-step feature description framework for 3D shape classification, and we discuss in detail its main algorithmic steps. Experimental results are presented in Section 4. Finally, we conclude in Section 5 and point out some future work directions.
\section{Background}
A 3D shape is usually modeled as a triangle mesh $\mathbb{M}$ whose vertices are sampled from a Riemannian manifold. A triangle mesh $\mathbb{M}$ may be defined as a graph $\mathbb{G}=(\mathcal{V},\mathcal{E})$ or $\mathbb{G}=(\mathcal{V},\mathcal{T})$, where $\mathcal{V}=\{\bm{v}_{1},\ldots,\bm{v}_{m}\}$ is the set of vertices, $\mathcal{E}=\{e_{ij}\}$ is the set of edges, and $\mathcal{T}$ is the set of triangles. Each edge $e_{ij}=[\bm{v}_{i},\bm{v}_{j}]$ connects a pair of vertices $\{\bm{v}_{i},\bm{v}_{j}\}$. Two distinct vertices $\bm{v}_{i},\bm{v}_{j}\in\mathcal{V}$ are adjacent (denoted by $\bm{v}_{i}\sim\bm{v}_{j}$ or simply $i\sim j$) if they are connected by an edge, i.e. $e_{ij}\in\mathcal{E}$.

\subsection{Laplace-Beltrami Operator}
Given a compact Riemannian manifold $\mathbb{M}$, the space $L^{2}(\mathbb{M})$ of all smooth, square-integrable functions on $\mathbb{M}$ is a Hilbert space endowed with inner product
$\langle f_{1},f_{2} \rangle=\int_{\mathbb{M}} f_{1}(\bm{x}) f_{2}(\bm{x})\,da(\bm{x})$, for all $f_{1}, f_{2}\in L^{2}(\mathbb{M})$, where $da(x)$ (or simply $dx$) denotes the measure from the area element of a Riemannian metric on $\mathbb{M}$. Given a twice-differentiable, real-valued function $f:\,\mathbb{M}\to\mathbb{R}$, the Laplace-Beltrami operator (LBO) is defined as $\Delta_{\mathbb{M}} f=-\mathrm{div}(\nabla_{\mathbb{M}} f)$, where $\nabla_{\mathbb{M}} f$ is the intrinsic gradient vector field and $\mathrm{div}$ is the divergence operator~\cite{Rosenberg:97}. The LBO is a linear, positive semi-definite operator acting on the space of real-valued functions defined on $\mathbb{M}$, and it is a generalization of the Laplace operator to non-Euclidean spaces.

\medskip
\noindent{\textbf{Discretization.}}\quad A real-valued function $f:\mathcal{V}\to\mathbb{R}$ defined on the mesh vertex set may be represented as an $m$-dimensional vector $\bm{f}=(f(i))\in\mathbb{R}^{m}$, where the $i$th component $f(i)$ denotes the function value at the $i$th vertex in $\mathcal{V}$. Using a mixed finite element/finite volume method on triangle meshes~\cite{Meyer:03}, the value of $\Delta_{\mathbb{M}}f$ at a vertex $\bm{v}_{i}$ (or simply $i$) can be approximated using the \textrm{cotangent weight} scheme as follows:
\begin{equation}
\Delta_{\mathbb{M}}f(i)\approx\frac{1}{a_{i}}\sum_{j\sim i}
\frac{\cot\alpha_{ij} + \cot\beta_{ij}}{2}\bigl(f(i)-f(j)\bigr),
\label{Eq:LapBeltrOperator}
\end{equation}
where $\alpha_{ij}$ and $\beta_{ij}$ are the angles $\angle(\bm{v}_{i}\bm{v}_{k_1}\bm{v}_{j})$ and $\angle(\bm{v}_{i}\bm{v}_{k_2}\bm{v}_{j})$ of two triangles $\{\bm{v}_{i},\bm{v}_{j},\bm{v}_{k_1}\}$ and $\{\bm{v}_{i},\bm{v}_{j},\bm{v}_{k_2}\}$ that are adjacent to the edge $[i,j]$, and $a_i$ is the area of the Voronoi cell at vertex $i$. It should be noted that the cotangent weight scheme is numerically consistent and preserves several important properties of the continuous LBO, including symmetry and positive semi-definiteness~\cite{Wardetzky:07}.

\medskip
\noindent{\textbf{Spectral Analysis.}}\quad The $m\times m$ matrix associated to the discrete approximation of the LBO is given by $\bm{L}=\bm{A}^{-1}\bm{W}$, where $\bm{A}=\mathrm{diag}(a_{i})$ is a positive definite diagonal matrix (mass matrix), and $\bm{W}=\diag(\sum_{k\neq i} c_{ik})-(c_{ij})$ is a sparse symmetric matrix (stiffness matrix). Each diagonal element $a_i$ is the area of the Voronoi cell at vertex $i$, and the weights $c_{ij}$ are given by
\begin{equation}
c_{ij}=
\begin{cases}
\dfrac{\cot\alpha_{ij} + \cot\beta_{ij}}{2} &\mbox{if } i\sim j \\
0 & \mbox{o.w.}
\end{cases}
\end{equation}
where $\alpha_{ij}$ and $\beta_{ij}$ are the opposite angles of two triangles that are adjacent to the edge $[i,j]$.

The eigenvalues and eigenvectors of $\bm{L}$ can be found by solving the generalized eigenvalue problem $\bm{W}\bg{\varphi}_{\ell}=\lambda_{\ell}\bm{A}\bg{\varphi}_{\ell}$ using for instance the Arnoldi method of ARPACK\footnote{ARPACK (ARnoldi PACKage) is a MATLAB library for computing the eigenvalues and eigenvectors of large matrices.}, where $\lambda_{\ell}$ are the eigenvalues and $\bg{\varphi}_{\ell}$ are the unknown associated eigenfunctions (i.e. eigenvectors which can be thought of as functions on the mesh vertices). We may sort the eigenvalues in ascending order as $0=\lambda_1<\lambda_{2}\le\dots\le\lambda_{m}$ with associated orthonormal eigenfunctions $\bg{\varphi}_{1}, \bg{\varphi}_{2},\dots,\bg{\varphi}_{m}$, where the orthogonality of the eigenfunctions is defined in terms of the $\bm{A}$-inner product, i.e.
\begin{equation}
\langle\bg{\varphi}_{k},\bg{\varphi}_{\ell}\rangle_{\bm{A}}=
\sum_{i=1}^{m}a_{i}\varphi_{k}(i)\varphi_{\ell}(i)=\delta_{k\ell},
\label{eq:orthoeig}
\end{equation}
for all $k, \ell=1,\dots,m$. We may rewrite the generalized eigenvalue problem in matrix form as $\bm{W}\bm{\Phi}=\bm{A}\bm{\Phi}\bm{\Lambda}$, where $\bm{\Lambda}=\diag(\lambda_{1},\dots,\lambda_{m})$ is an $m\times m$ diagonal matrix with the $\lambda_{\ell}$ on the diagonal, and $\bm{\Phi}$ is an $m\times m$ orthogonal matrix whose $\ell$th column is the unit-norm eigenvector $\bg{\varphi}_{\ell}$.

\subsection{Spectral Shape Signatures}
In recent years, several local descriptors based on the eigensystem of the LBO have been proposed in the 3D shape analysis literature, including the heat kernel signature (HKS) and wave kernel signature (WKS)~\cite{Sun:09,Aubry:11}. Both HKS and WKS have an elegant physical interpretation: the HKS describes the amount of heat remaining at a mesh vertex $j\in\mathcal{V}$ after a certain time, whereas the WKS is the probability of measuring a quantum particle with the initial energy distribution at $j$. The HKS at a vertex $j$ is defined as:
\begin{equation}
\mathfrak{s}_{t_k}(j)=\sum_{\ell=1}^{m}e^{-\lambda_{\ell}t_{k}}\varphi_{\ell}^{2}(j),
\end{equation}
where $\lambda_{\ell}$ and $\varphi_{\ell}$ are the eigenvalues and eigenfunctions of the LBO.

The HKS contains information mainly from low frequencies, which correspond to macroscopic features of the shape; and thus exhibits a major discrimination ability in shape retrieval tasks. With multiple scaling factors $t_k$, a collection of low-pass filters are established. The larger is $t_k$, the more high frequencies are suppressed. However, different frequencies are always mixed in the HKS, and high-precision localization tasks may fail due in part to the suppression of the high frequency information, which corresponds to microscopic features. To circumvent these disadvantages, Aubry \textit{et al.}~\cite{Aubry:11} introduced the WKS, which is defined at a vertex $j$ as follows:
\begin{equation}
\mathfrak{s}_{t_k}(j)=\sum_{\ell=1}^{m}C_{t_k}\exp\left (-\frac{(\log{t_k}-\log{\lambda_{\ell}})^2}{\sigma^2} \right )\varphi_{\ell}^{2}(j),
\end{equation}
where $C_{t_k}$ is a normalization constant. The WKS explicitly separates the influences of different frequencies, treating all frequencies equally. Thus, different spatial scales are naturally separated, making the high-precision feature localization possible.

\medskip
Given a range of discrete scales $t_k$, a bank of filters is constructed for each signature, and thus a vertex $j$ on the mesh surface can be described by a $p$-dimensional point signature vector given by
\begin{equation}
\bm{s}_{j} = \{\mathfrak{s}_{t_k}(j)\mid k=1,\dots,p \},\quad\text{for } j=1,\dots,m.
\end{equation}

\section{Method}
In this section, we provide a detailed description of our proposed 3D shape classification method that utilizes spectral graph wavelets in conjunction with the BoF paradigm. Shape classification is the process of organizing a dataset of shapes into a known number of classes, and the task is to assign new shapes to one of these classes. It is common practice in classification to randomly split the available data into training and test sets. Classification aims to learn a classifier (also called predictor or classification model) from labeled training data. The training data consist of a set of training examples or instances that are labeled with predefined classes. The resulting, trained model is subsequently applied to the test data to classify future (unseen) data instances into these classes. The test data, which consists of data instances with unknown class labels, is used to evaluate the performance of the classification model and determine its accuracy in terms of the number of test instances correctly or incorrectly predicted by the model. A good classifier should result in high accuracy, or equivalently, in few misclassifications.

In our proposed framework, each 3D shape in the dataset is first represented by local descriptors, which are arranged into a spectral graph wavelet signature matrix. Then, we perform soft-assignment coding by embedding local descriptors into the visual vocabulary space, resulting in mid-level features which we refer to as spectral graph wavelet codes (SGWC). It is important to point out that the vocabulary is computed offline by concatenating all the spectral graph wavelet signature matrices into a data matrix, followed by applying the K-means algorithm to find the data cluster centers.

\begin{figure*}[htb]
\centering
\includegraphics[scale=.8]{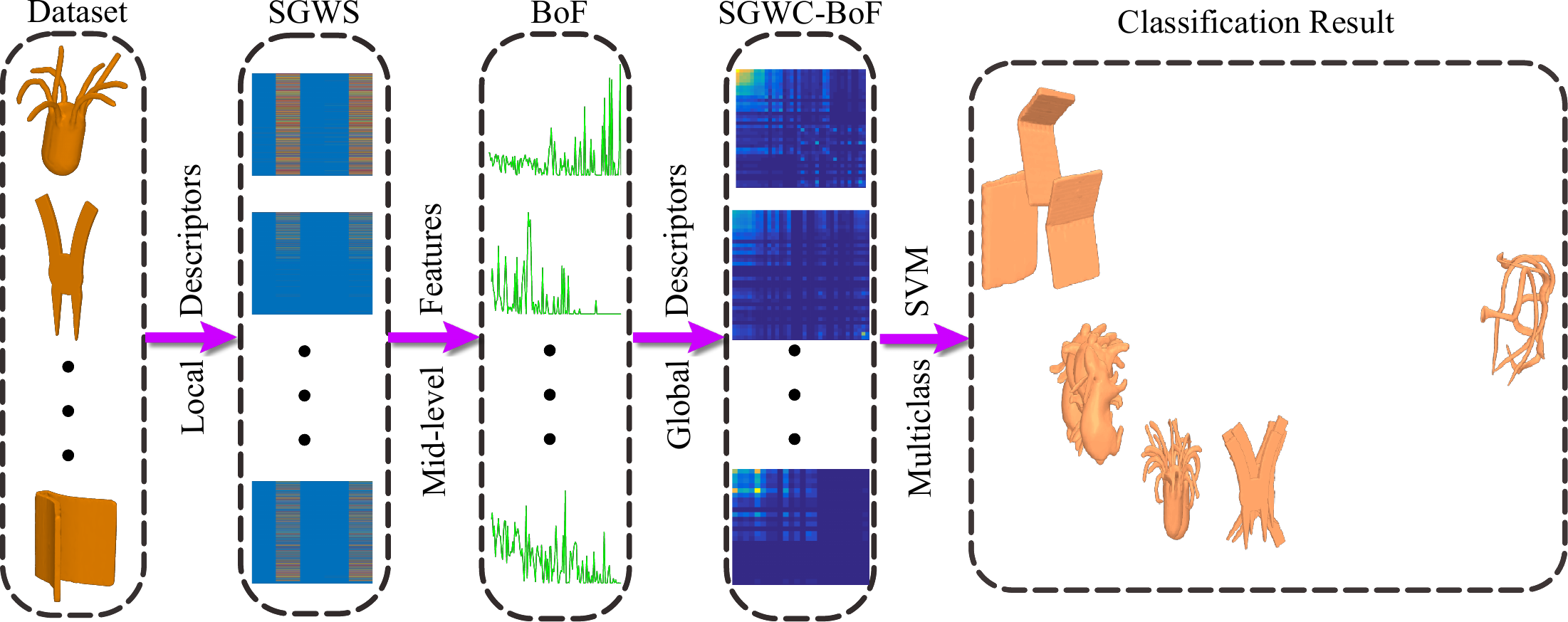}
\caption{Flowchart of the proposed approach.}
\label{Fig:flowchart}
\end{figure*}

In a bid to capture the spatial relations between features, we compute a global descriptor of each shape in terms of a geodesic exponential kernel and mid-level features, resulting in a SGWC-BoF matrix which is then transformed into a SGWC-BoF vector by stacking its columns one underneath the other. The last stage of the proposed approach is to perform classification on the SGWC-BoF vectors using a classification algorithm. The flowchart of the proposed framework is depicted in Figure~\ref{Fig:flowchart}. Multiclass support vector machines (SVMs) are widely used supervised learning methods for classification. Supervised learning algorithms consist of two main steps: training step and test step. In the training step, a classification model (classifier) is learned from the training data by a learning algorithm (e.g., SVMs). In the test step, the learned model is evaluated using a set of test data to predict the class labels for the classifier and hence assess the classification accuracy.

\subsection{Spectral Graph Wavelet Transform}
For any graph signal $f: \mathcal{V}\to\mathbb{M}$, the forward and inverse graph Fourier transforms (also called manifold harmonic and inverse manifold harmonic transforms) are defined as
\begin{equation}
\hat{f}(\ell)=\langle \bm{f},\bg{\varphi}_{\ell}\rangle=\sum_{i=1}^{m} a_{i} f(i)\varphi_{\ell}(i),\quad \ell=1,\dots,m
\label{eq:MHT}
\end{equation}
and
\begin{equation}
f(i)=\sum_{\ell=1}^{m} \hat{f}(\ell)\varphi_{\ell}(i)=\sum_{\ell=1}^{m} \langle \bm{f},\bg{\varphi}_{\ell}\rangle\varphi_{\ell}(i),\quad i\in\mathcal{V},
\label{eq:IMHT}
\end{equation}
respectively, where $\hat{f}(\ell)$ is the value of $f$ at eigenvalue $\lambda_\ell$ (i.e. $\hat{f}(\ell)=\hat{f}(\lambda_\ell)$). In particular, the graph Fourier transform of a delta function $\delta_j$ centered at vertex $j$ is given by
\begin{equation}
\hat{\delta_{j}}(\ell)=\sum_{i=1}^{m} a_{i}\delta_{j}(i)\varphi_{\ell}(i)=
\sum_{i=1}^{m} a_{i}\delta_{ij}\varphi_{\ell}(i)=a_{j}\varphi_{\ell}(j).
\end{equation}
The forward and inverse graph Fourier transforms may be expressed in matrix-vector multiplication as follows:
\begin{equation}
\hat{\bm{f}}=\bm{\Phi}^{\T}\bm{A}\bm{f}\quad\text{and}\quad \bm{f}=\bm{\Phi}\hat{\bm{f}},
\end{equation}
where $\bm{f}=(f(i))$ and $\hat{\bm{f}}=(\hat{f}(\ell))$ are $m$-dimensional vectors whose elements are given by \eqref{eq:MHT} and \eqref{eq:IMHT}, respectively.

\medskip
\noindent{\textbf{Wavelet Function.}}\quad The spectral graph wavelet transform is determined by the choice of a spectral graph wavelet generating kernel $g:\mathbb{R}^+\rightarrow \mathbb{R}^+$, which is analogous to the Fourier domain wavelet. To act as a band-pass filter, the kernel $g$ should satisfy $g(0)=0$ and $\lim_{x\rightarrow \infty }g(x)=0$.

Let $g$ be a given kernel function and denote by $T_g^t$ the wavelet operator at scale $t$. Similar to the Fourier domain, the graph Fourier transform of $T_g^t$ is given by
\begin{equation}
\widehat{T_g^tf}(\ell)=g(t\lambda_\ell)\hat{f}(\ell),
\label{Eq:Tgf}
\end{equation}
where $g$ acts as a scaled band-pass filter. Thus, the inverse graph Fourier transform is given by
\begin{equation}
(T_{g}^{t} f)(i)=\sum_{\ell=1}^{m}\widehat{T_g^tf}(\ell)\varphi_{\ell}(i)=
\sum_{\ell=1}^{m}g(t\lambda_{\ell})\hat{f}(\ell)\varphi_{\ell}(i).
\label{SGWcoef}
\end{equation}
Applying the wavelet operator $T_{g}^{t}$ to a delta function centered at vertex $j$ (i.e. $f(i)=\delta_{j}(i)=\delta_{ij}$), the spectral graph wavelet $\psi_{t,j}$ localized at vertex $j$ and scale $t$ is then given by
\begin{equation}
\begin{split}
\psi_{t,j}(i)=(T_{g}^{t}\delta_j)(i) &= \sum_{\ell=1}^{m}g(t\lambda_{\ell})\hat{\delta_j}(\ell)\varphi_{\ell}(i) \\
&=
\sum_{\ell=1}^{m} a_{j}g(t\lambda_\ell)\varphi_{\ell}(j)\varphi_{\ell}(i).
\end{split}
\label{SGWpsi}
\end{equation}
This indicates that shifting the wavelet to vertex $j$ corresponds to a multiplication by $\varphi_{\ell}(j)$. It should be noted that $g(t\lambda_{\ell})$ is able to modulate the spectral wavelets $\psi_{t,j}$ only for $\lambda_{\ell}$ within the domain of the spectrum of the LBO. Thus, an upper bound on the largest eigenvalue $\lambda_{\max}$ is required to provide knowledge on the spectrum in practical applications.

Hence, the spectral graph wavelet coefficients of a given function $f$ can be generated from its inner product with the spectral graph wavelets:
\begin{equation}
W_f(t,j)= \langle \bm{f},\bg{\psi}_{t,j} \rangle=\sum_{\ell=1}^{m} a_{j}g(t\lambda_{\ell})\hat{f}(\ell)\varphi_{\ell}(j).
\label{Eq:SGWcoefficients}
\end{equation}

\medskip
\noindent{\textbf{Scaling Function.}}\quad Similar to the low-pass scaling functions in the classical wavelet analysis, a second class of waveforms $h:\mathbb{R}^+\rightarrow \mathbb{R}$ are used as low-pass filters to better encode the low-frequency content of a function $f$ defined on the mesh vertices. To act as a low-pass filter, the scaling function $h$ should satisfy $h(0)>0$ and $h(x)\rightarrow0$ as $x\rightarrow\infty$. Similar to the wavelet kernels, the scaling functions are given by
\begin{equation}
\begin{split}
\phi_{j}(i)=(T_{h}\delta_j)(i) &= \sum_{\ell=1}^{m}h(\lambda_{\ell})\hat{\delta_j}(\ell)\varphi_{\ell}(i) \\
&=\sum_{\ell=1}^{m} a_{j} h(\lambda_\ell)\varphi_{\ell}(j)\varphi_{\ell}(i).
\end{split}
\end{equation}
and their spectral coefficients are
\begin{equation}
S_f(j)=\langle \bm{f},\bg{\phi}_{j} \rangle=\sum_{\ell=1}^{m} a_{j} h(\lambda_{\ell})\hat{f}(\ell)\varphi_{\ell}(j).
\end{equation}
A major advantage of using the scaling function is to ensure that the original signal $f$ can be stably recovered when sampling scale parameter $t$ with a discrete number of values $t_k$. As demonstrated in~\cite{Hammond:11}, given a set of scales $\{t_k\}_{k=1}^{K}$, the set $F=\{\phi_j\}_{j=1}^{m}\cup \{\psi_{t_k,j}\}_{k=1\;j=1}^{K\;\;\;\;\; m}$ forms a spectral graph wavelet frame with bounds
\begin{equation}
A=\min_{\lambda\in [0,\lambda_{\max}]} G(\lambda)\quad\text{and}\quad B=\max_{\lambda\in [0,\lambda_{\max}]} G(\lambda),
\end{equation}
where
\begin{equation}
G(\lambda)=h(\lambda)^2+\sum_k g(t_k \lambda)^2.
\end{equation}
The stable recovery of $f$ is ensured when $A$ and $B$ are away from zero. Additionally, the crux of the scaling function is to smoothly represent the low-frequency content of the signal on the mesh. Thus, the design of the scaling function $h$ is uncoupled from the choice of the wavelet generating kernel $g$.

\subsection{Local Descriptors}
Wavelets are useful in describing functions at different levels of resolution. To characterize the localized context around a mesh vertex $j\in\mathcal{V}$, we assume that the signal on the mesh is a unit impulse function, that is $f(i)=\delta_j(i)$ at each mesh vertex $i\in\mathcal{V}$. Thus, it follows from~\eqref{SGWcoef} that the spectral graph wavelet coefficients are
\begin{equation}
W_{\delta_j}(t,j)=\langle \bg{\delta}_{j},\bg{\psi}_{t,j} \rangle=\sum_{\ell=1}^{m}a_{j}^{2}g(t\lambda_\ell)\varphi_{\ell}^{2}(j),
\label{DeltaW_coefficients}
\end{equation}
and that the coefficients of the scaling function are
\begin{equation}
S_{\delta_j}(j)=\sum_{\ell=1}^{m}a_{j}^{2} h(\lambda_\ell)\varphi_{\ell}^{2}(j).
 \label{DeltaS_coefficients}
\end{equation}
Following the multiresolution analysis, the spectral graph wavelet and scaling function coefficients are collected to form the spectral graph wavelet signature at vertex $j$ as follows:
\begin{equation}
\bm{s}_j=\{\bm{s}_{L}(j)\mid L=1,\dots,R\},
 \label{Eq:SGWSignature}
\end{equation}
where $R$ is the resolution parameter, and $\bm{s}_{L}(j)$ is the shape signature at resolution level $L$ given by
\begin{equation}
\bm{s}_{L}(j)=\{W_{\delta_j}(t_k,j)\mid k=1,\dots,L\}\cup\{S_{\delta_j}(j)\}.
 \label{Eq:SGWSignatureLevel}
\end{equation}
The wavelet scales $t_k$ ($t_k > t_{k+1}$) are selected to be logarithmically equispaced between maximum and minimum scales $t_1$ and $t_L$, respectively. Thus, the resolution level $L$ determines the resolution of scales to modulate the spectrum. At resolution $R=1$, the spectral graph wavelet signature $\bm{s}_j$ is a 2-dimensional vector consisting of two elements: one element, $W_{\delta_j}(t_1,j)$, of spectral graph wavelet function coefficients and another element, $S_{\delta_j}(j)$, of scaling function coefficients. And at resolution $R=2$, the spectral graph wavelet signature $\bm{s}_j$ is a 5-dimensional vector consisting of five elements (four elements of spectral graph wavelet function coefficients and one element of scaling function coefficients). In general, the dimension of a spectral graph wavelet signature $\bm{s}_j$ at vertex $j$ can be expressed in terms of the resolution $R$ as follows:
\begin{equation}
p = \frac{(R+1)(R+2)}{2}-1.
\end{equation}
Hence, for a $p$-dimensional signature $\bm{s}_j$, we define a $p\times m$ spectral graph wavelet signature matrix as $\bm{S}=(\bm{s}_1,\dots,\bm{s}_m)$, where $\bm{s}_j$ is the signature at vertex $j$ and $m$ is the number of mesh vertices. In our implementation, we used the Mexican hat wavelet as a kernel generating function $g$. In addition, we used the scaling function $h$ given by
\begin{equation}
h(x)=\gamma \exp\left(-\left (\frac{x}{ 0.6\lambda_{\min}} \right)^4\right),
\end{equation}
where $\lambda_{\min}=\lambda_{\max}/20$ and $\gamma$ is set such that $h(0)$ has the same value as the maximum value of $g$. The maximum and minimum scales are set to $t_{1}=2/\lambda_{\min}$ and $t_{L}=2/\lambda_{\max}$.

\begin{figure*}[!htb]
\centering
\begin{tabular}{cc}
\includegraphics[width=7.3cm,height=4.2cm]{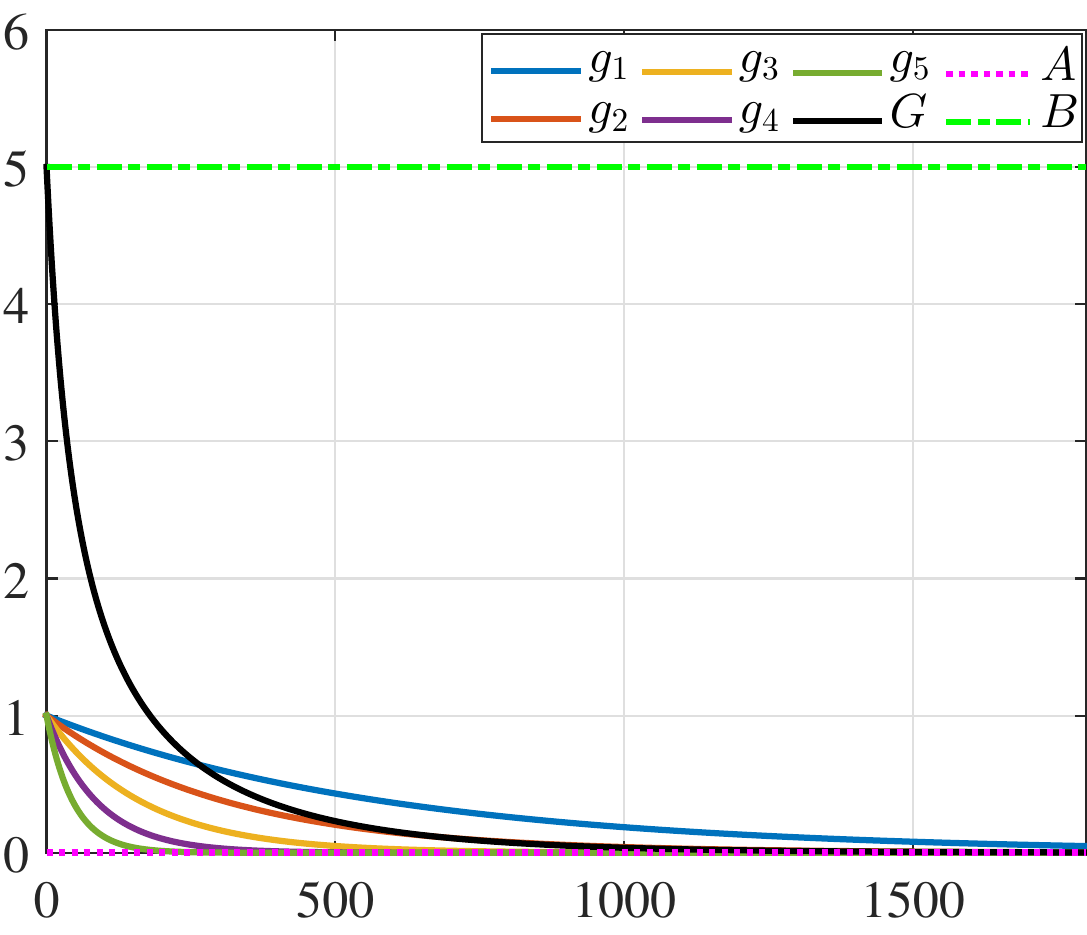}&
\includegraphics[width=7.3cm,height=4.2cm]{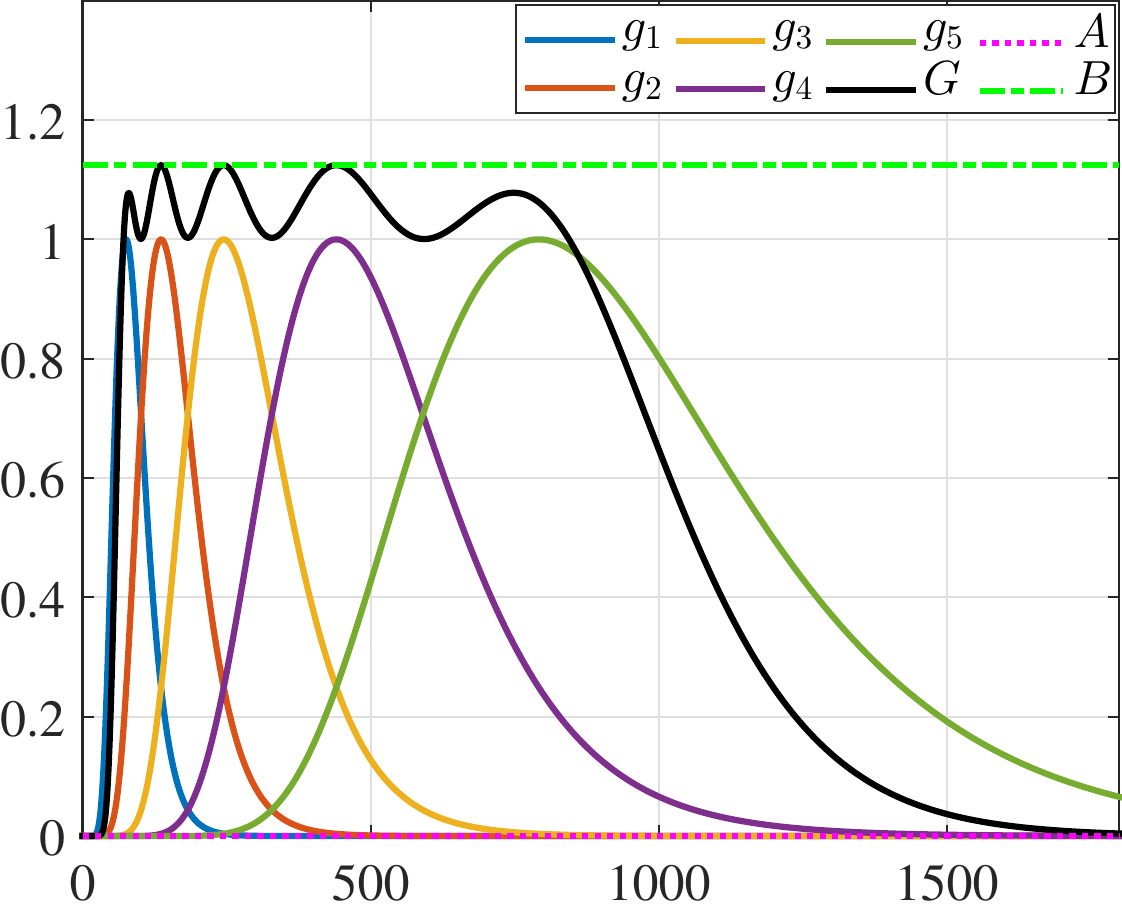}\\
(a) Heat kernel &  (b) Wave kernel\\
\includegraphics[width=7.5cm,height=4.2cm]{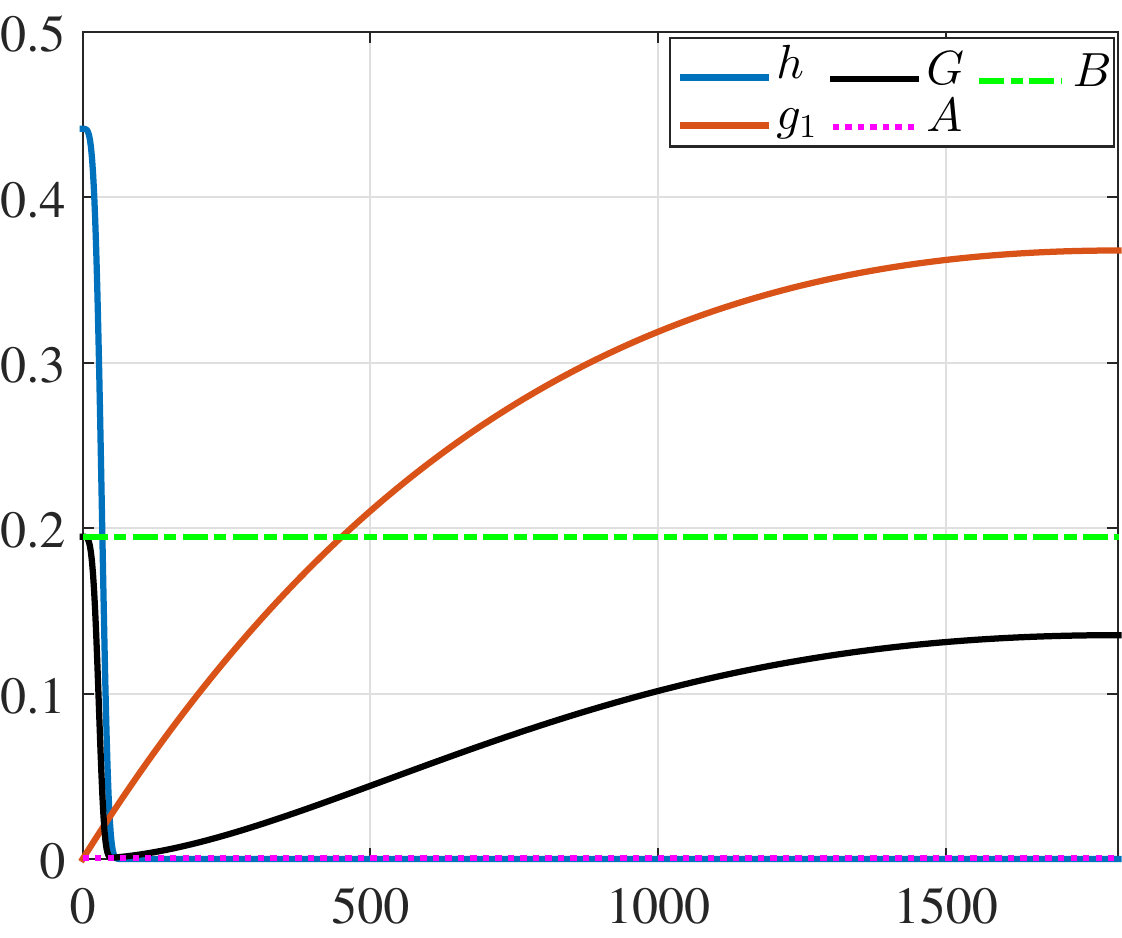}&
\includegraphics[width=7.5cm,height=4.2cm]{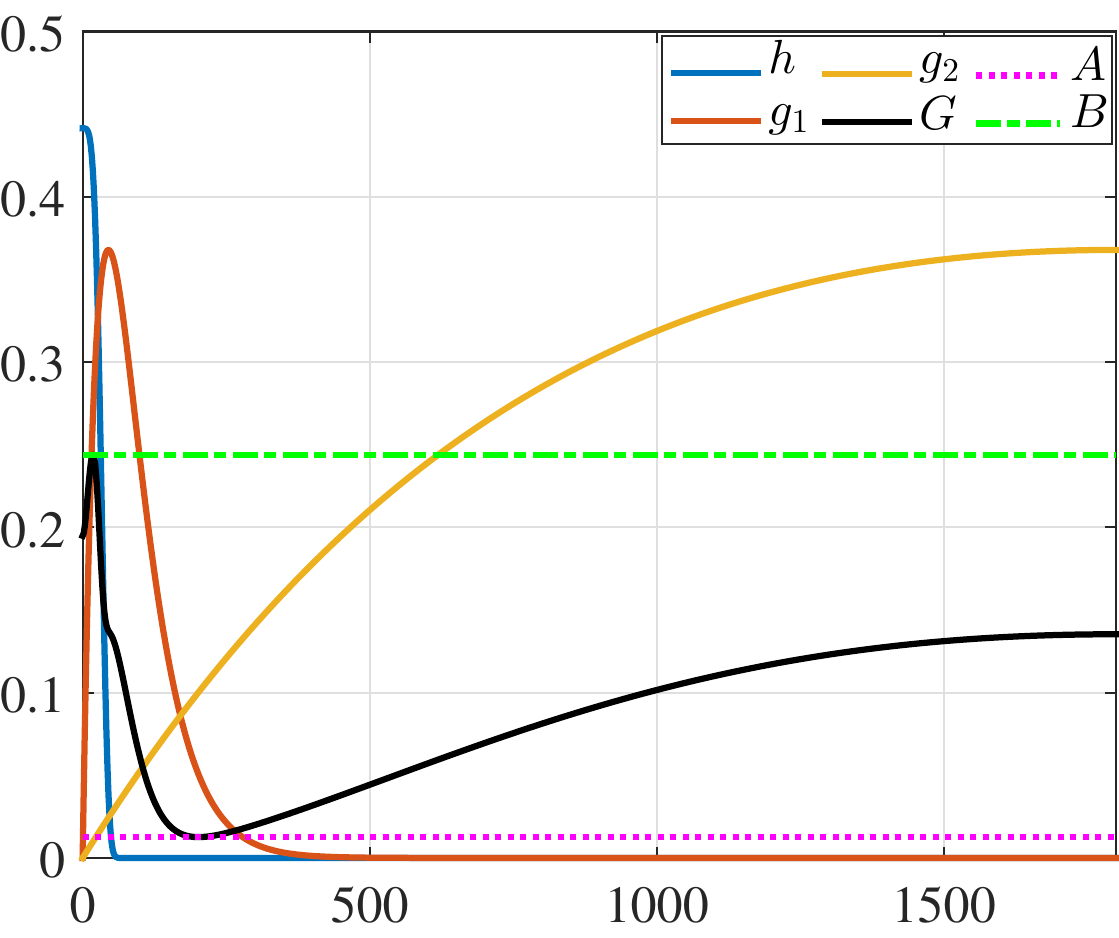}\\
(c) Mexican hat kernel for \emph{R}=1 & (d) Mexican hat kernel for \emph{R}=2\\
\includegraphics[width=7.5cm,height=4.2cm]{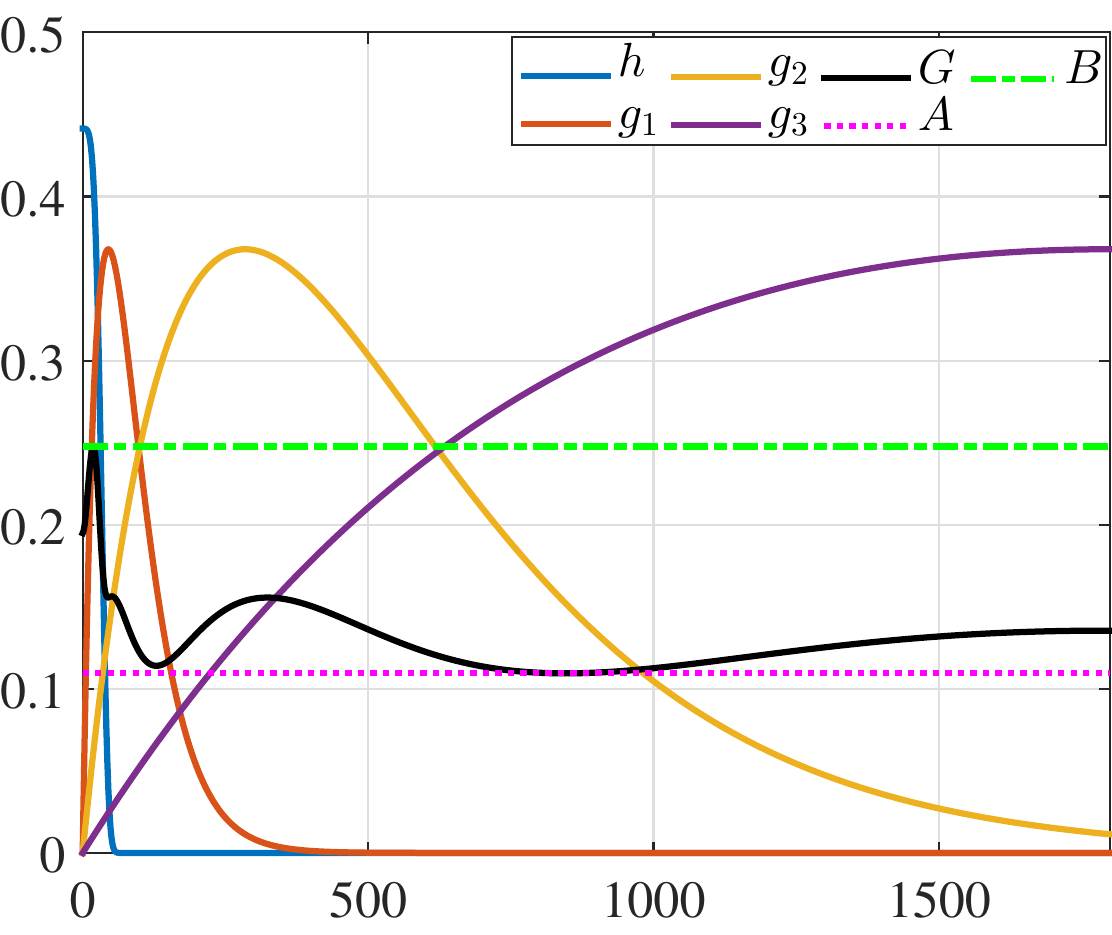}&
\includegraphics[width=7.5cm,height=4.2cm]{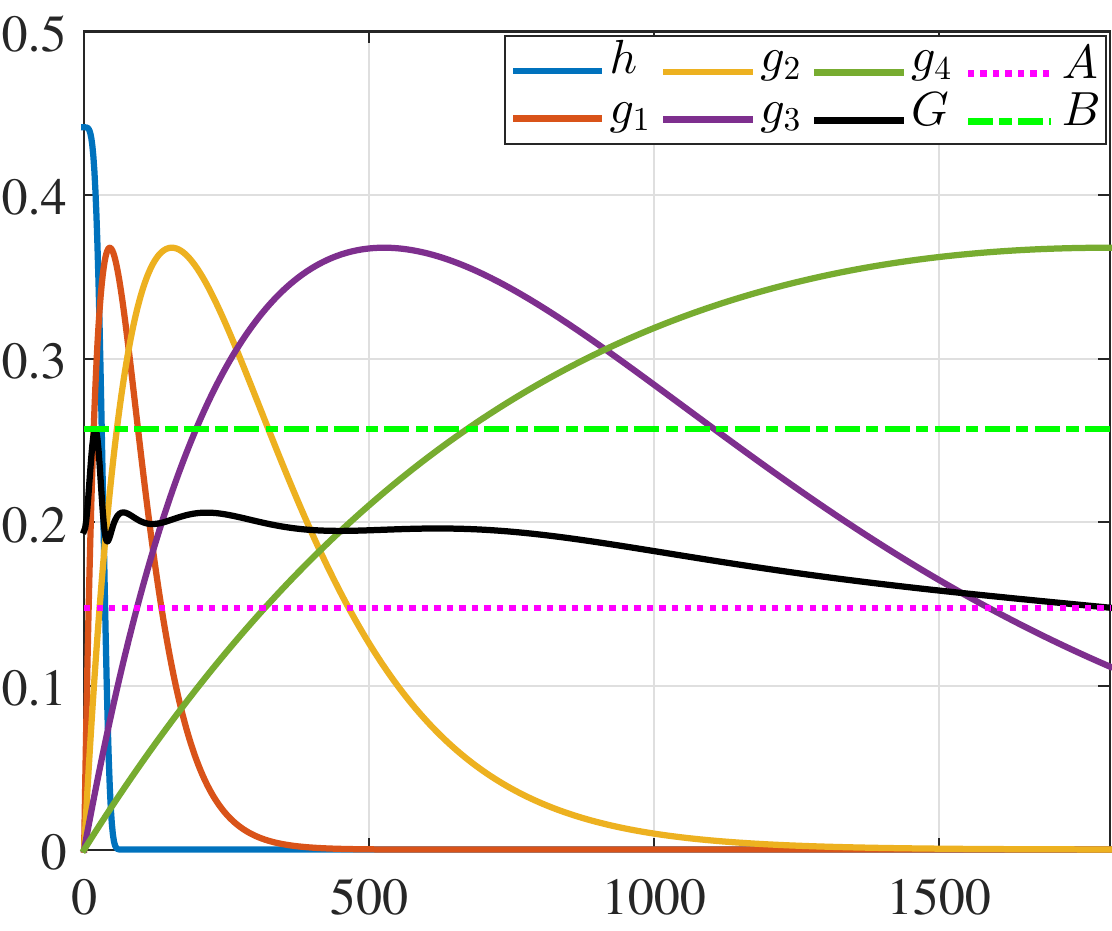}\\
(e) Mexican hat kernel for \emph{R}=4 & (f) Mexican hat kernel for \emph{R}=4\\
\includegraphics[width=7.5cm,height=4.2cm]{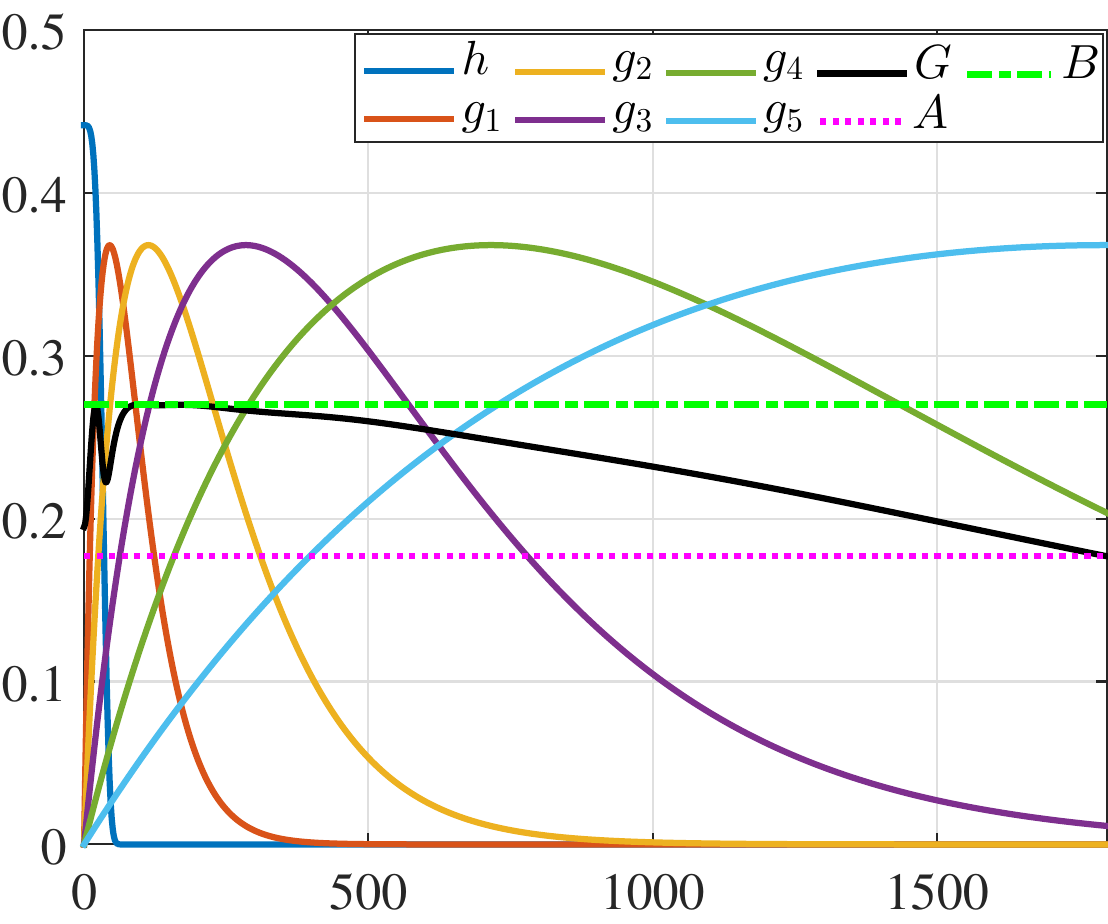}&
\includegraphics[width=7.5cm,height=4.2cm]{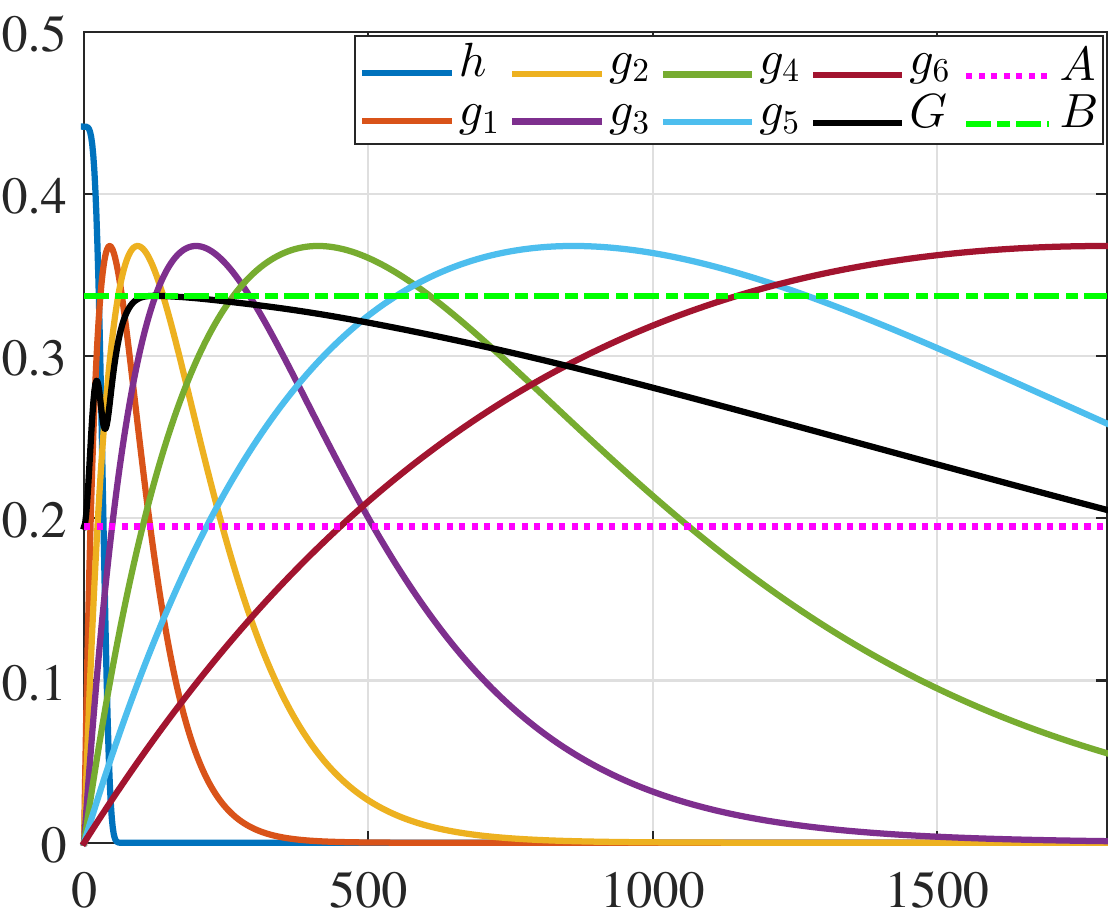}\\
(g) Mexican hat kernel for \emph{R}=5 & (h) Mexican hat kernel for \emph{R}=6\\
\end{tabular}
\caption{Spectrum modulation using different kernel functions at various resolutions. The dark line is the squared sum function $G$, while the dash-dotted and the dotted lines are upper and lower bounds ($B$ and $A$) of $G$, respectively.}
\label{Fig:SGWT}
\end{figure*}

The geometry captured at each resolution $R$ of the spectral graph wavelet signature can be viewed as the area under the curve $G$ shown in Figure~\ref{Fig:SGWT}. For a given resolution $R$, we can understand the information from a specific range of the spectrum as its associated areas under $G$. As the resolution $R$ increases, the partition of spectrum becomes tighter, and thus a larger portion of the spectrum is highly weighted.

\subsection{Mid-Level Features}
\begin{figure*}[!htb]
\setlength{\tabcolsep}{.1em}
\centering
\begin{tabular}{cccc}
\includegraphics[scale=0.12]{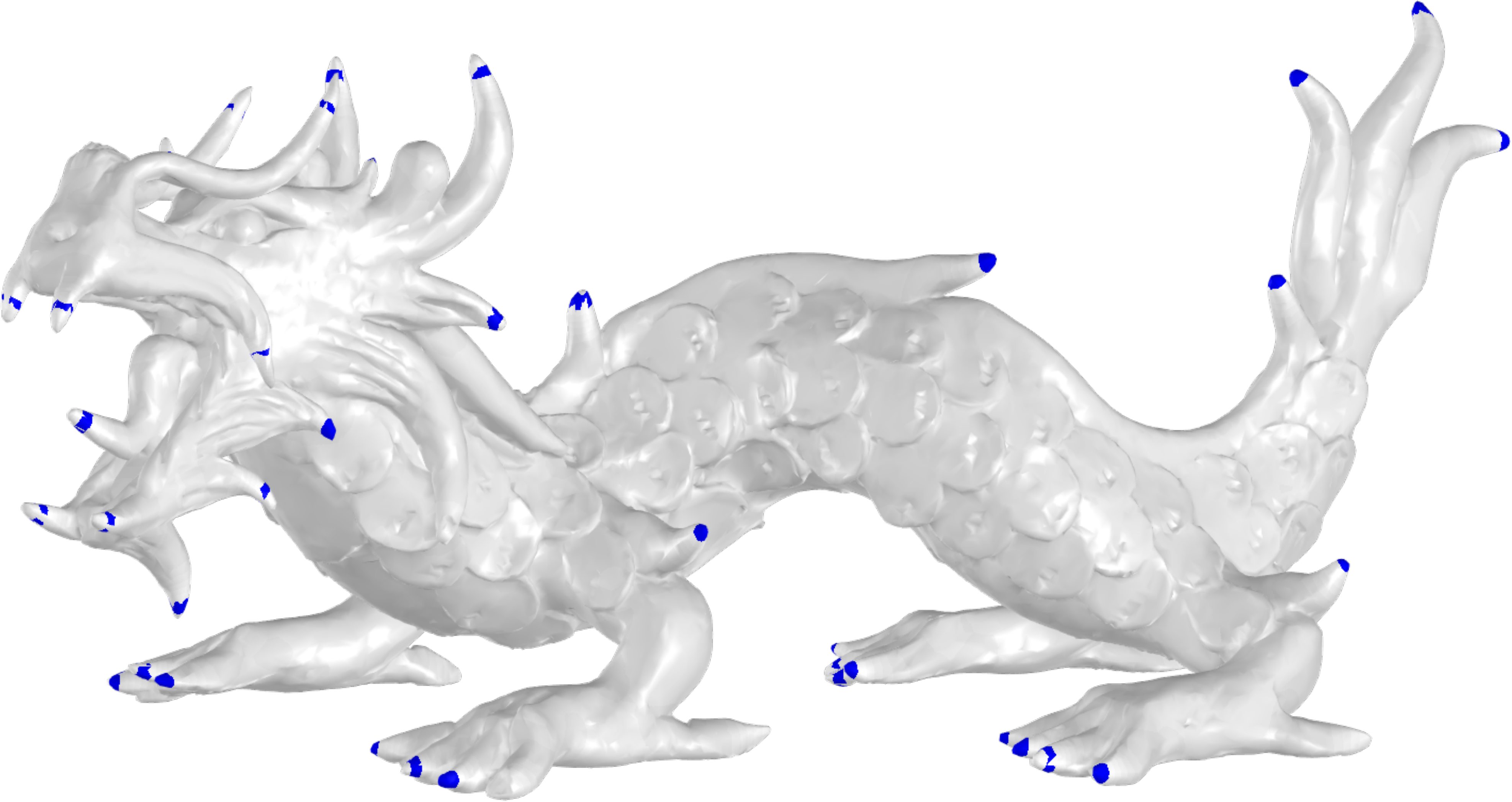}&
\includegraphics[scale=0.12]{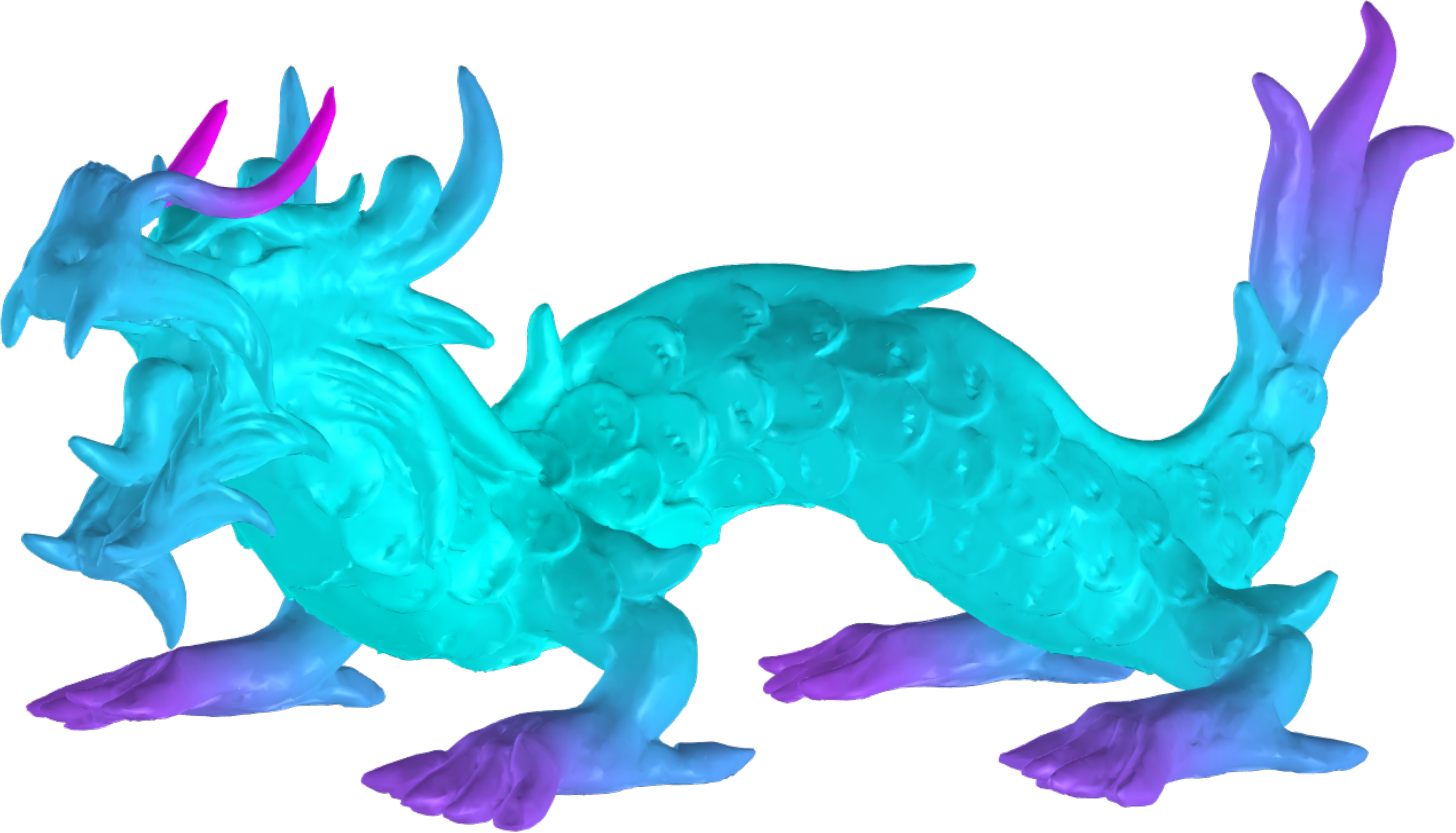} &
\includegraphics[scale=0.12]{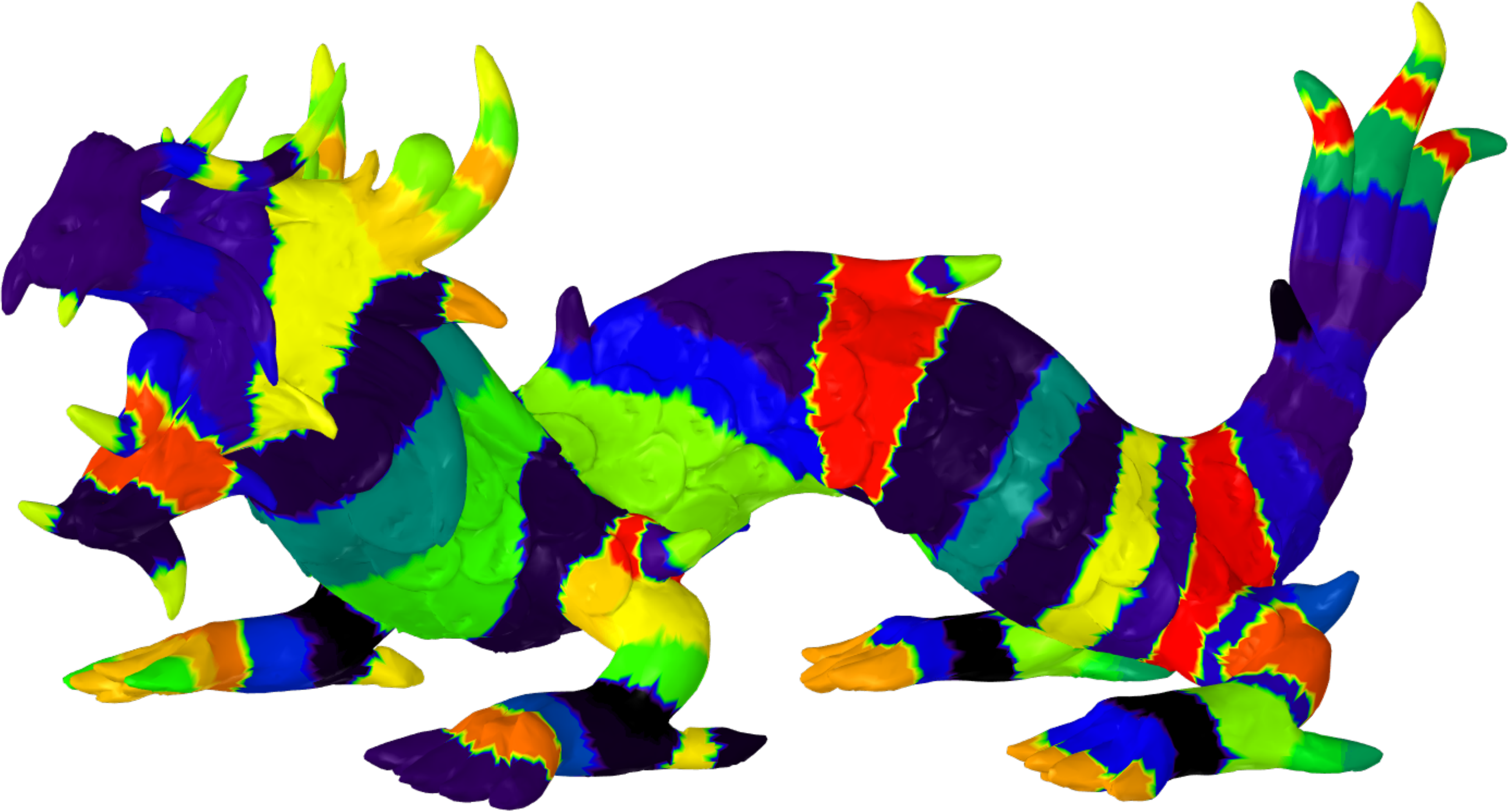}&
\includegraphics[scale=0.16]{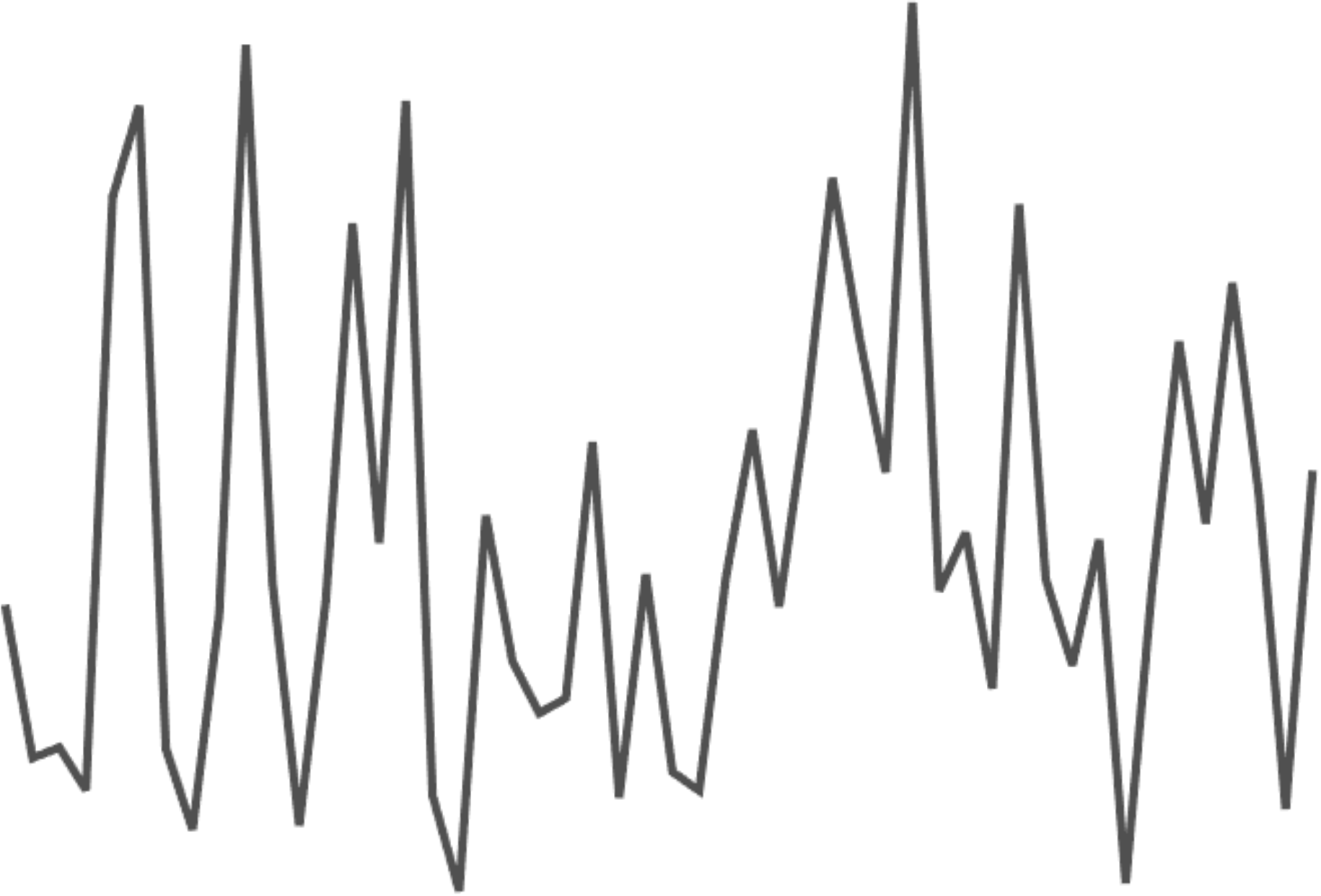} \\
Feature extraction &  Feature description & Vector quantization & Bag of features
\end{tabular}
\caption{Flow of the bag-of-features model.}
\label{BOF}
\end{figure*}

The BoF model aggregates local descriptors of a shape in an effort to provide a simple representation that may be used to facilitate comparison between shapes. We model each 3D shape as a triangle mesh $\mathbb{M}$ with $m$ vertices. The BoF model consists of four main steps: feature extraction and description, codebook design, feature coding and feature pooling. The idea of the BoF paradigm on 3D shapes is illustrated in Figure~\ref{BOF}.

\medskip
\noindent\textbf{Feature Extraction and Description.}\quad In the BoF paradigm, a 3D shape $\mathbb{M}$ is represented as a collection of $m$ local descriptors of the same dimension $p$, where the order of different feature vectors is of no importance. Local descriptors may be classified into two main categories: dense and sparse. Dense descriptors are computed at each point (vertex) of the shape, while sparse descriptors are computed by identifying a set of salient points using a feature detection algorithm. In our proposed framework, we represent $\mathbb{M}$ by a $p\times m$ matrix $\bm{S}=(\bm{s}_{1},\dots,\bm{s}_{m})$ of spectral graph wavelet signatures, where each $p$-dimensional feature vector $\bm{s}_{i}$ is a dense, local descriptor that encodes the local structure around the $i$th vertex of $\mathbb{M}$.

\medskip
\noindent\textbf{Codebook Design.}\quad A codebook (or visual vocabulary) is constructed via clustering by quantizing the $m$ local descriptors (i.e. spectral graph wavelet signatures) into a certain number of codewords. These codewords are usually defined as the centers $\bm{v}_{1},\dots,\bm{v}_{k}$ of $k$ clusters obtained by performing an unsupervised learning algorithm (e.g., vector quantization via K-means clustering) on the signature matrix $\bm{S}$. The codebook is the set $\mathcal{V}=\{\bm{v}_{1},\dots,\bm{v}_{k}\}$ of size $k$, which may be represented by a $p\times k$ vocabulary matrix $\bm{V}=(\bm{v}_{1},\dots,\bm{v}_{k})$.

\medskip
\noindent\textbf{Feature Coding.}\quad The goal of feature coding is to embed local descriptors in the vocabulary space. Each spectral graph wavelet signature $\bm{s}_i$ is mapped to a codeword $\bm{v}_r$ in the codebook via the $k\times m$ cluster soft-assignment matrix $\bm{U}=(u_{ri})=(\bm{u}_{1},\dots,\bm{u}_{m})$ whose elements are given by
\begin{equation}
 u_{ri}=\dfrac{\exp(-\alpha\lVert\bm{s}_{i} - \bm{v}_{r}\rVert_{2}^{2})}
  {\sum_{\ell=1}^{k}\exp(-\alpha\lVert\bm{s}_{i} - \bm{v}_{\ell}\rVert_{2}^{2})},
\end{equation}
where $\lVert\cdot\rVert_{2}$ denotes the $L_{2}$-norm, and $\alpha$ is a smoothing parameter that controls the softness of the assignment. Unlike hard-assignment coding in which a local descriptor is assigned to the nearest cluster, soft-assignment coding assigns descriptors to every cluster center with different probabilities in an effort to improve quantization properties of the coding step. We refer to the coefficient vector $\bm{u}_i$ as the \textit{spectral graph wavelet code} (SGWC) of the descriptor $\bm{s}_i$, with $u_{ri}$ being the coefficient with respect to the codeword $\bm{v}_r$.

\medskip
\noindent\textbf{Histogram Representation (Feature Pooling).}\quad Each spectral graph wavelet signature is mapped to a certain codeword through the clustering process and the shape is then represented by the histogram $\bm{h}$ of the codewords, which is a $k$-dimensional vector given by
\begin{equation}
\bm{h}=\bm{U}\bm{1}_{m}=(h_{r})_{r=1,\dots,k},
\end{equation}
where $h_{r}=\sum_{i=1}^{m}u_{ri}$. That is, the histogram consists of the column-sums of the cluster assignment matrix $\bm{U}$. Other feature pooling methods include average- and max-pooling. In general, any predefined pooling function that aggregates the information of different codewords into a single feature vector can be used.

\subsection{Global Descriptors}
A major drawback of the BoF model is that it only considers the distribution of the codewords and disregards all information about the spatial relations between features, and hence the descriptive ability and discriminative power of the BoF paradigm may be negatively impacted. To circumvent this limitation, various solutions have been recently proposed including the spatially sensitive bags of features (SS-BoF)~\cite{Bronstein:11} and geodesic-aware bags of features (GA-BoF)~\cite{Bu:14}. The SS-BoF, which is defined in terms of mid-level features and the heat kernel, can be represented by a square matrix whose elements represent the frequency of appearance of nearby codewords in the vocabulary. Similarly, the GA-BoF matrix is obtained by replacing the heat kernel in the SS-BoF with a geodesic exponential kernel. Unlike the heat kernel which is time-dependent, the geodesic exponential kernel avoids the possible effect of time scale and shape size~\cite{Bu:14}. In the same vein, we define a global descriptor of a shape as a $k\times k$ SGWC-BoF matrix $\bm{F}$ defined in terms of spectral graph wavelet codes and a geodesic exponential kernel as follows:
\begin{equation}
\bm{F} = \bm{U}\bm{K}\bm{U}^{\T},
\end{equation}
where $\bm{U}$ is a $k\times m$ matrix of spectral graph wavelet codes (i.e. mid-level features), and $\bm{K}=(\kappa_{ij})$ is an $m\times m$ geodesic exponential kernel matrix whose elements are given by
\begin{equation}
\kappa_{ij} = \exp\left(-\frac{d_{ij}}{\epsilon}\right),
\end{equation}
with $d_{ij}$ denoting the geodesic distance between any pair of mesh vertices $\bm{v}_i$ and $\bm{v}_j$, and $\epsilon$ is a positive, carefully chosen parameter that determines the width of the kernel. Intuitively, the parameter $\epsilon$ controls the linearity of the kernel function, i.e. the larger the width, the linear the function. It is worth pointing out that the proposed SGWC-BoF is similar in spirit to SS-BoF and GA-BoF. The main distinction of our work is that we use multiresolution local descriptors that may be regarded as generalized signatures for those in ~\cite{Bronstein:11,Bu:14}. In addition, our spectral graph wavelet signature combines the advantages of both band-pass and low-pass filters.
\subsection{Multiclass Support Vector Machines}
SVMs are supervised learning models that have proven effective in solving classification problems. SVMs are based upon the idea of maximizing the margin, i.e. maximizing the minimum distance from the separating hyperplane to the nearest example. Although SVMs were originally designed for binary classification, several extensions have been proposed in the literature to handle the multiclass classification. The idea of multiclass SVM is to decompose the multiclass problem into multiple binary classification tasks that can be solved efficiently using binary SVM classifiers. One of the simplest and most widely used coding designs for multiclass classification is the one-vs-all approach, which constructs $K$ binary SVM classifiers such that for each binary classifier, one class is positive and the rest are negative. In other words, the one-vs-all approach requires $K$ binary SVM classifiers, where the $i$th classifier is trained with positive examples belonging to class $i$ and negative examples belonging to the remaining $K-1$ classes. When testing an unknown example, the classifier producing the maximum output (i.e. largest value of the decision function) is considered the winner, and this class label is assigned to that example.

\subsection{Proposed Algorithm}
Shape classification is a supervised learning method that assigns shapes in a dataset to target classes. The objective of 3D shape classification is to accurately predict the target class for each 3D shape in the dataset. Our proposed 3D shape classification algorithm consists of four main steps. The first step is to represent each 3D shape in the dataset by a spectral graph wavelet signature matrix, which is a feature matrix consisting of local descriptors. More specifically, let $\mathcal{D}$ be a dataset of $n$ shapes modeled by triangle meshes $\mathbb{M}_1,\dots,\mathbb{M}_n$. We represent each 3D shape in the dataset $\mathcal{D}$ by a $p\times m$ spectral graph wavelet signature matrix $\bm{S}=(\bm{s}_{1},\dots,\bm{s}_{m})\in\mathbb{R}^{p\times m}$, where $\bm{s}_i$ is the $p$-dimensional local descriptor at vertex $i$ and $m$ is the number of mesh vertices.

In the second step, the spectral graph wavelet signatures $\bm{s}_{i}$ are mapped to high-dimensional mid-level feature vectors using the soft-assignment coding step of the BoF model, resulting in a $k\times m$ matrix $\bm{U}=(\bm{u}_{1},\dots,\bm{u}_{m})$ whose columns are the $k$-dimensional mid-level feature codes (i.e. SGWC). In the third step, the $k\times k$ SGWC-BoF matrix $\bm{F}$ is computed using the mid-level feature codes matrix and a geodesic exponential kernel, followed by reshaping $\bm{F}$ into a $k^2$-dimensional global descriptor $\bm{x}_i$. In the fourth step, the SGWC-BoF vectors $\bm{x}_i$ of all $n$ shapes in the dataset are arranged into a $k^{2}\times n$ data matrix $\bm{X}=(\bm{x}_{1},\dots,\bm{x}_{n})$. Finally, a one-vs-all multiclass SVM classifier is performed on the data matrix $\bm{X}$ to find the best hyperplane that separates all data points of one class from those of the other classes.

The task in multiclass classification is to assign a class label to each input example. More precisely, given a training data of the form $\mathcal{X}_{\textrm{train}}=\{(\bm{x}_{i},y_{i})\}$, where $\bm{x}_{i}\in\mathbb{R}^{k^{2}}$ is the $i$th example (i.e. SGWC-BoF vector) and $y_{i}\in\{1,\dots,K\}$ is its $i$th class label, we aim at finding a learning model that contains the optimized parameters from the SVM algorithm. Then, the trained SVM model is applied to a test data $\mathcal{X}_{\textrm{test}}$, resulting in predicted labels $\hat{y}_{i}$ of new data. These predicted labels are subsequently compared to the labels of the test data to evaluate the classification accuracy of the model.

To assess the performance of the proposed framework, we employed two commonly-used evaluation criteria, the confusion matrix and accuracy, which will be discussed in more detail in the next section. The main algorithmic steps of our approach are summarized in Algorithm 1.
\begin{algorithm}
  \caption{Spectral graph wavelet classifier}\label{algo:blah}
  \begin{algorithmic}[1]
    \REQUIRE Dataset $\mathcal{D}=\{\mathbb{M}_1,\dots,\mathbb{M}_n\}$ of $n$ 3D shapes
    \ENSURE $n$-dimensional vector $\hat{\bm{y}}$ containing predicted class labels for each 3D shape
    \FOR{$i=1$ to $n$}
    \STATE Compute the $p\times m$ spectral graph wavelet signature matrix $\bm{S}_{i}$ of each shape $\mathbb{M}_{i}$
    \STATE Apply soft-assignment coding to find the $k\times m$ mid-level feature matrix $\bm{U}_{i}$, where $k>p$
    \STATE Compute the $k\times k$ SGWC-BoF matrix $\bm{F}_{i}$, and reshape it into a $k^2$-dimensional vector $\bm{x}_{i}$
    \ENDFOR
    \STATE Arrange all the $n$ SGWC-BoF vectors into a $k^{2}\times n$ data matrix $\bm{X}=(\bm{x}_1,\dots,\bm{x}_n)$
    \STATE Perform multiclass SVM on $\bm{X}$ to find the $n$-dimensional vector $\hat{\bm{y}}$ of predicted class labels.
  \end{algorithmic}
\end{algorithm}

\noindent It is important to point out that in our implementation the vocabulary is computed offline by applying the K-means algorithm to the $p\times mn$ matrix obtained by concatenating all SGWS matrices of all $n$ meshes in the dataset. As a result, the vocabulary is a matrix $\bm{V}$ of size $p\times k$, where $k>p$.

\section{Experiments}
In this section, we conduct extensive experiments to evaluate the performance of the proposed SGWC-BoF framework for 3D shape classification. The effectiveness of our approach is validated by performing a comprehensive comparison with several state-of-the-art methods.

\begin{figure}[!htb]
\centering
\includegraphics[width=3.4in,height=2.1in]{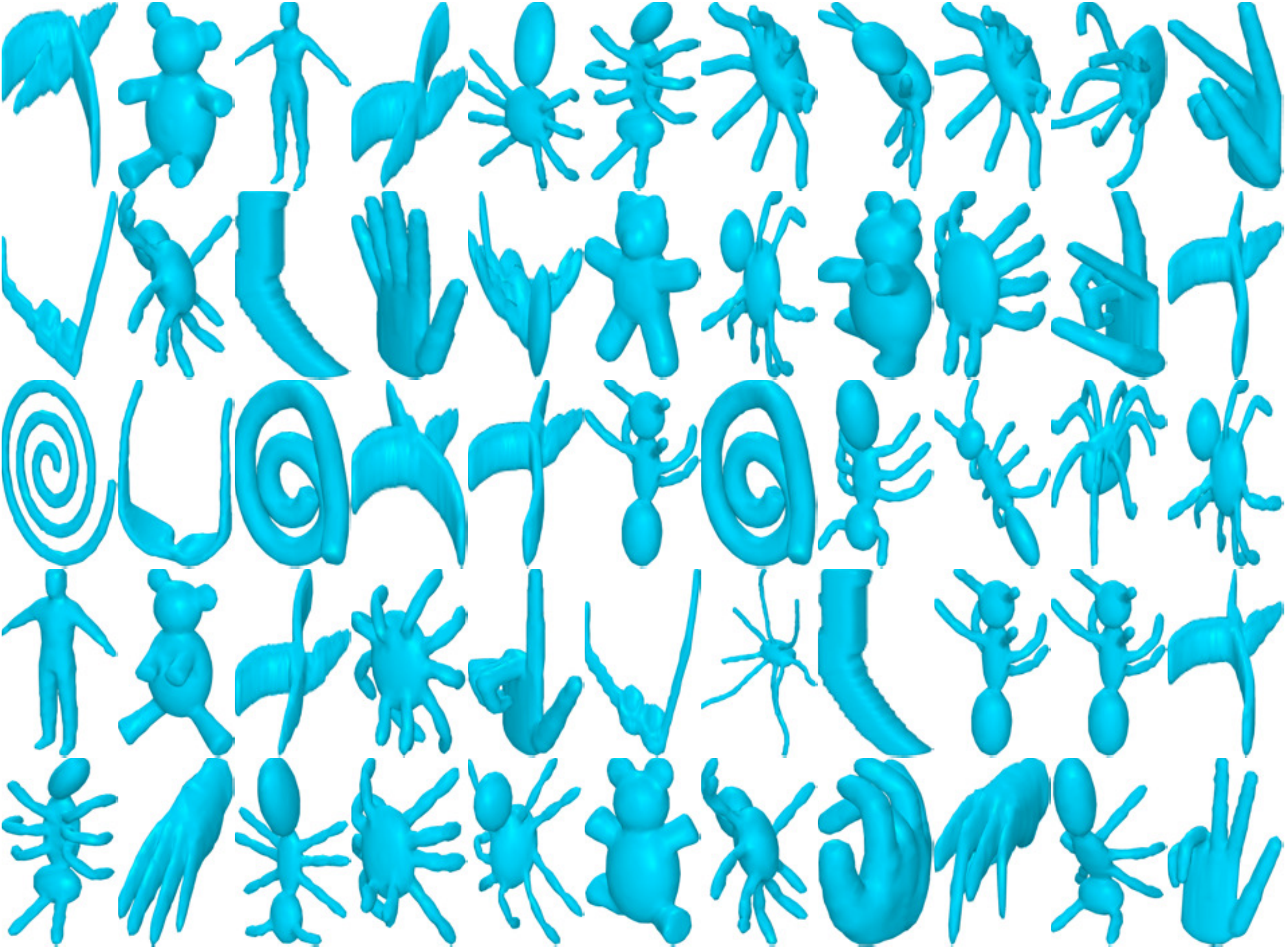} \\ [3ex]
\includegraphics[width=3.4in,height=2.1in]{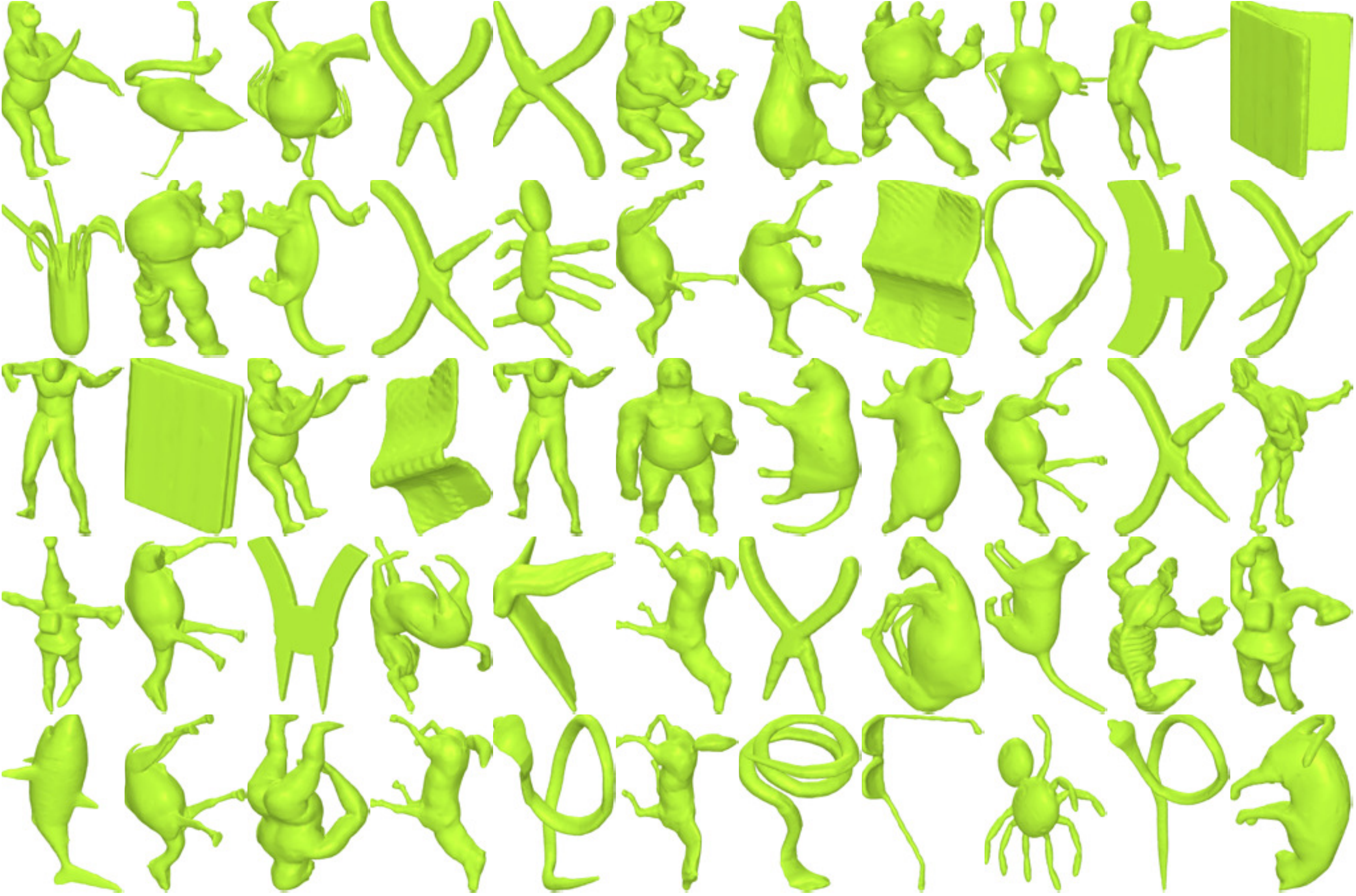}
\caption{Sample shapes from SHREC-2010 (top) and SHREC-2011 (bottom). }
\label{Fig:shrec1011}
\end{figure}

\medskip
\noindent{\textbf{Datasets.}}\quad The performance of the proposed framework is evaluated on two standard and publicly available 3D shape benchmarks: SHREC-2010 and SHREC-2011. Sample shapes from these two benchmarks are shown in Figure~\ref{Fig:shrec1011}.

\medskip
\noindent{\textbf{Performance Evaluation Measures.}}\quad In practice, the available data (which has classes) $\mathcal{X}$ for classification is usually split into two disjoint subsets: the training set $\mathcal{X}_{\textrm{train}}$ for learning, and the test set $\mathcal{X}_{\textrm{test}}$ for testing. The training and test sets are usually selected by randomly sampling a set of training instances from $\mathcal{X}$ for learning and using the rest of instances for testing. The performance of a classifier is then assessed by applying it to test data with known target values and comparing the predicted values with the known values. One important way of evaluating the performance of a classifier is to compute its confusion matrix (also called contingency table), which is a $K\times K$ matrix that displays the number of correct and incorrect predictions made by the classifier compared with the actual classifications in the test set, where $K$ is the number of classes.

Another intuitively appealing measure is the classification accuracy, which is a summary statistic that can be easily computed from the confusion matrix as the total number of correctly classified instances (i.e. diagonal elements of confusion matrix) divided by the total number of test instances. Alternatively, the accuracy of a classification model on a test set may be defined as follows
\begin{equation}
\begin{split}
\text{Accuracy} &=\frac{\text{Number of correct classifications}}{\text{Total number of test cases}} \\
&= \frac{\lvert \bm{x}\,:\,\bm{x}\in\mathcal{X}_{\textrm{test}}\,\wedge\, \hat{y}(\bm{x}) = y(\bm{x})\rvert}{\lvert \bm{x}\,:\,\bm{x}\in\mathcal{X}_{\textrm{test}}\rvert},
\end{split}
\end{equation}
where $y(\bm{x})$ is the actual (true) label of $\bm{x}$, and $\hat{y}(\bm{x})$ is the label predicted by the classification algorithm. A correct classification means that the learned model predicts the same class as the original class of the test case. The error rate is equal to one minus accuracy.

\medskip
\noindent{\textbf{Baseline Methods.}}\quad For each of the 3D shape benchmarks used for experimentation, we will report the comparison results of our method against various state-of-the-art methods, including Shape-DNA~\cite{Reuter:06}, compact Shape-DNA~\cite{Gao:14}, GPS embedding~\cite{Chaudhari:14}, GA-BoF~\cite{Bu:14}, and F1-, F2-, and F3-features~\cite{Khabou:07}. The latter features, which are defined in terms of the Laplacian matrix eigenvalues, were shown to have good inter-class discrimination capabilities in 2D shape recognition~\cite{Gao:14}, but they can easily be extended to 3D shape analysis using the eigenvalues of the LBO.

\medskip
\noindent{\textbf{Implementation Details.}}\quad The experiments were conducted on a desktop computer with an Intel Core i5 processor running at 3.10 GHz and 8 GB RAM; and all the algorithms were implemented in MATLAB. The appropriate dimension (i.e. length or number of features) of a shape signature is problem-dependent and usually determined experimentally. For fair comparison, we used the same parameters that have been employed in the baseline methods, and in particular the dimensions of shape signatures. In our setup, a total of 201 eigenvalues and associated eigenfunctions of the LBO were computed. For the proposed approach, we set the resolution parameter to $R=2$ (i.e. the spectral graph wavelet signature matrix is of size $5\times m$, where $m$ is the number of mesh vertices) and the kernel width of the geodesic exponential kernel to $\epsilon=0.1$. Moreover, the parameter of the soft-assignment coding is computed as $\alpha=1/(8\mu^2)$, where $\mu$ is the median size of the clusters in the vocabulary~\cite{Bronstein:11}. For shape-DNA, GPS embedding, and F1-, F2-, and F3-features, the selected number of retained eigenvalues equals $10$. As suggested in~\cite{Gao:14}, the dimension of the compact Shape-DNA signature was set to $33$.

\subsection{SHREC-2010 Dataset}
SHREC-2010 is a dataset of 3D shapes consisting of 200 watertight mesh models from 10 classes~\cite{Lian:SHREC10}. These models are selected from the McGill Articulated Shape Benchmark dataset. Each class contains 20 objects with distinct postures. Moreover, each shape in the dataset has approximately $m=1002$ vertices.

\begin{figure}[htb]
\centering
\includegraphics[scale=0.72]{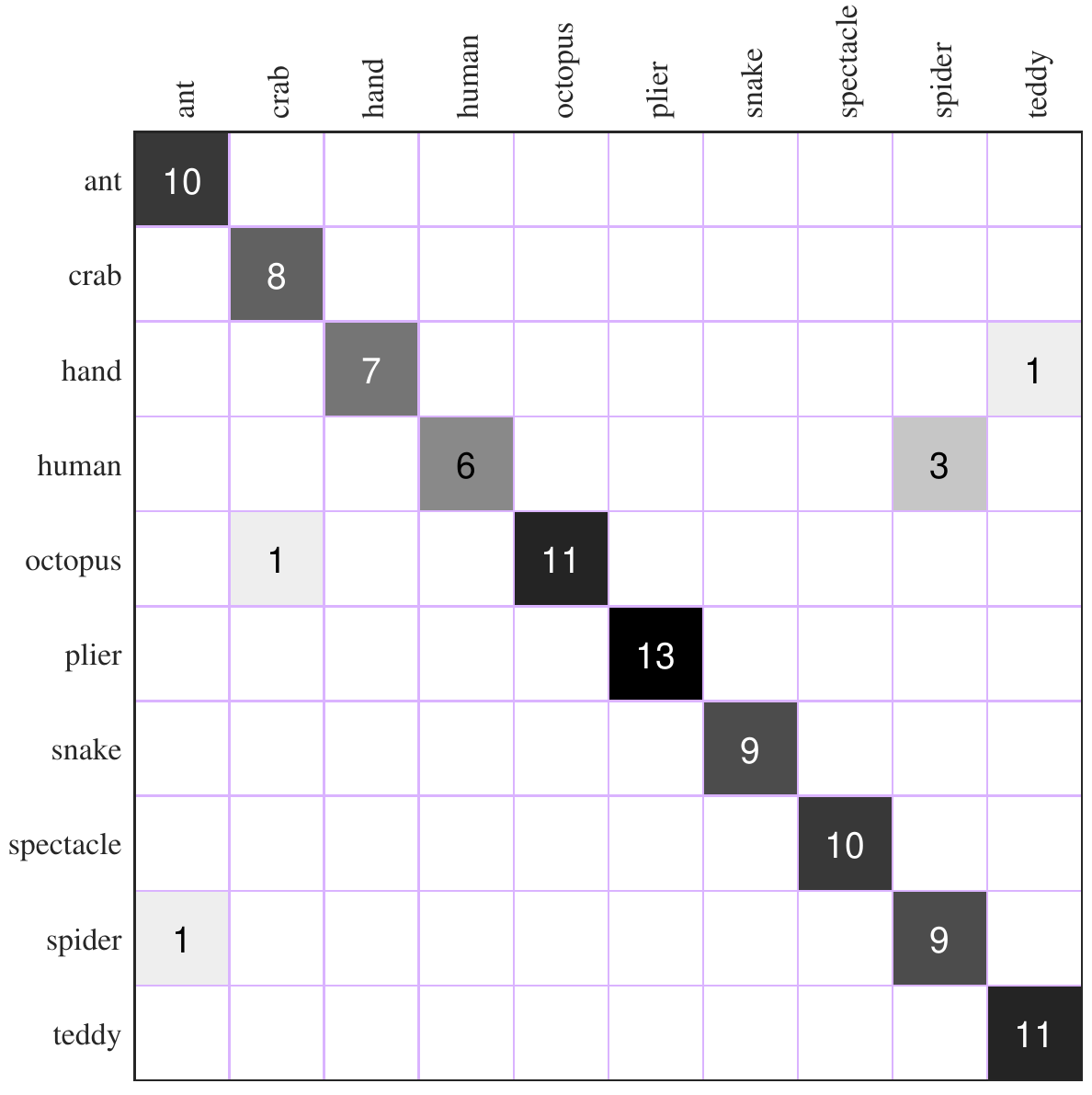}
\caption{Confusion matrix for SHREC-2010 using linear multiclass SVM.}
\label{Fig:confSHREC10}
\end{figure}

\medskip
\noindent{\textbf{Performance Evaluation.}}\quad We randomly selected 50\% shapes in the SHREC-2010 dataset to hold out for the test set, and the remaining shapes for training. That is, the test data consists of 100 shapes. A one-vs-all multiclass SVM is first trained on the training data to learn the model (i.e. classifier), which is subsequently used on the test data with known target values in order to predict the class labels.
Figure~\ref{Fig:confSHREC10} displays the confusion matrix for SHREC-2010 on the test data. This $10\times 10$ confusion matrix shows how the predictions are made by the model. Its rows correspond to the actual (true) class of the data (i.e. the labels in the data), while its columns correspond to the predicted class (i.e. predictions made by the model). The value of each element in the confusion matrix is the number of predictions made with the class corresponding to the column for instances with the correct value as represented by the row. Thus, the diagonal elements show the number of correct classifications made for each class, and the off-diagonal elements show the errors made. As can be seen in Figure~\ref{Fig:confSHREC10}, the proposed approach was able to accurately classify all shapes in the test data, except the hand, octopus and spider models which were misclassified only once as teddy, crab and ant, respectively. Also, the human shape was misclassified three times as a spider. Such a good performance strongly suggests that our method captures well the discriminative features of the shapes.

\begin{figure}[!htb]
\setlength{\tabcolsep}{.3em}
\centering
\begin{tabular}{cc}
\includegraphics[scale=.14]{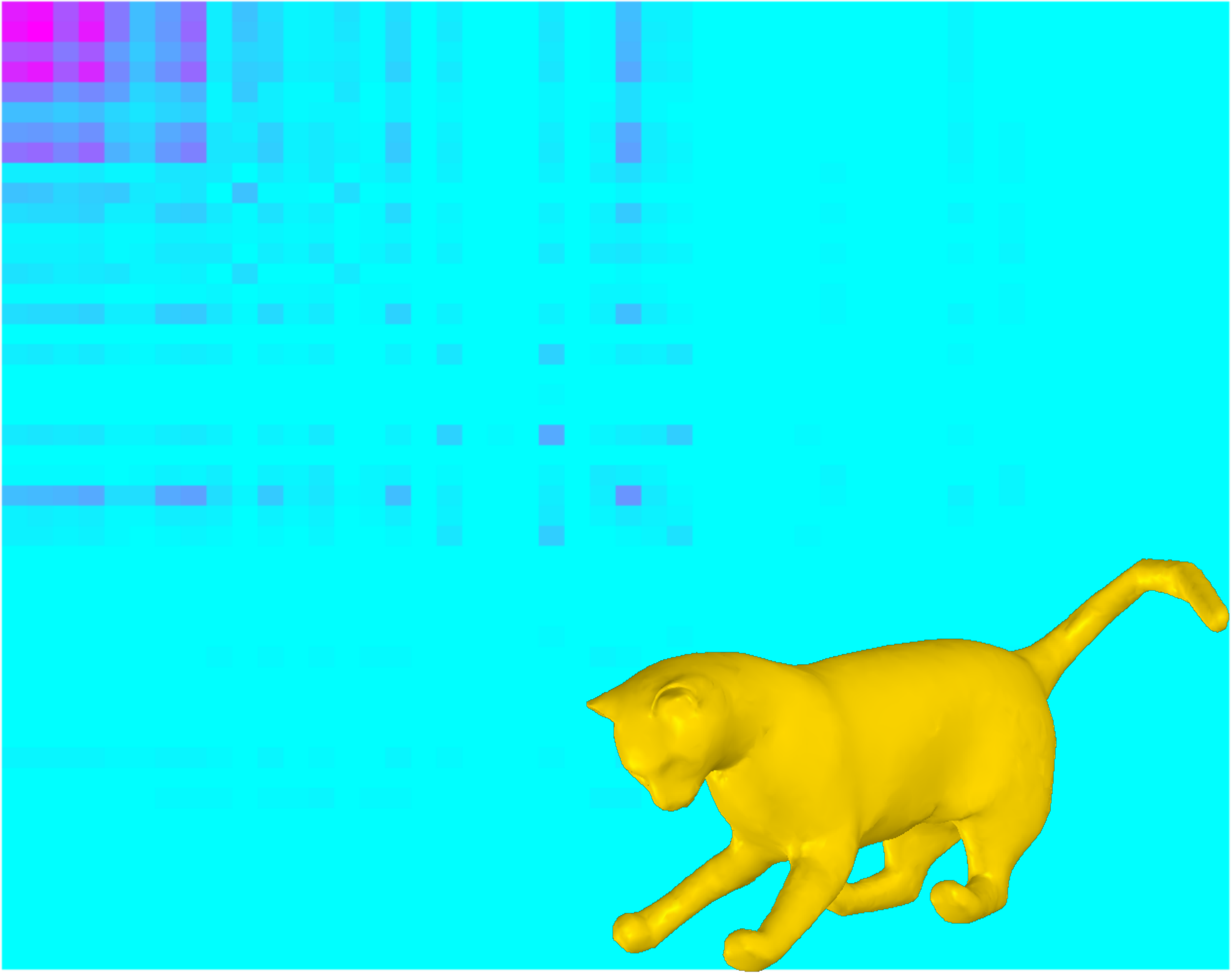} &
\includegraphics[scale=.14]{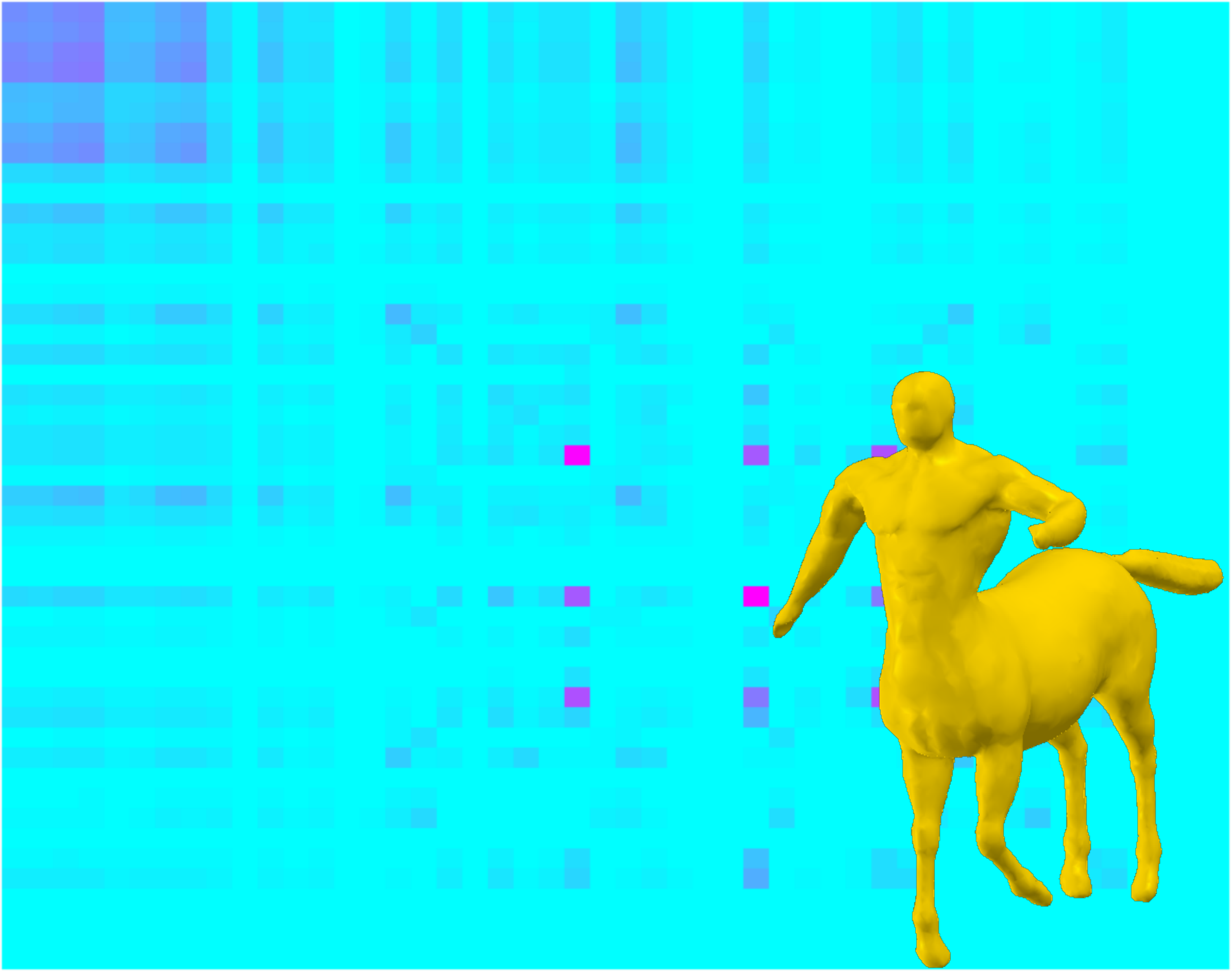}
\end{tabular}
\caption{Spectral graph wavelet codes of two shapes (cat and centaur) from two different classes of the SHREC-2010 dataset.}
\label{Fig:shrec10codes}
\end{figure}

\medskip
\noindent{\textbf{Results.}}\quad In our approach, each 3D shape in the SHREC-2010 dataset is represented by a $5\times 1002$ matrix of spectral graph wavelet signatures. Setting the number of codewords to $k=128$, we computed offline a $5\times 128$ vocabulary matrix $\bm{V}$ via K-means clustering. The pre-computation of the vocabulary of size 128 took approximately 15 minutes. The soft-assignment coding of the BoF model yields a $128\times 1002$ matrix $\bm{U}$ of spectral graph wavelet codes, resulting in a SGWC-BoF data matrix $\bm{X}$ of size $128^{2}\times 200$. Figure~\ref{Fig:shrec10codes} shows the spectral graph wavelet code matrices of two shapes from two different classes of SHREC-2010. As can be seen, the global descriptors are quite different and hence they may be used efficiently to discriminate between shapes in classification tasks.

We compared the proposed method to Shape-DNA, compact shape-DNA, GPS embedding, and F1-, F2-, and F3-features. In order to compute the accuracy, we repeated the experimental process 10 times with different randomly selected training and test data in an effort to obtain reliable results, and the accuracy for each run was recorded, then we selected the best result of each method. The classification accuracy results are summarized in Table~\ref{Tab:SHREC10class}, which shows the results of the baseline methods and the proposed framework. As can be seen, our SGWC-BOF method achieves better performance than Shape-DNA, compact Shape-DNA, GPS embedding, GA-BoF, and F1-, F2-, and F3-features. The proposed approach yields the highest classification accuracy of 95.66\%, with performance improvements of 2.76\% and 4.70\% over the best baseline methods cShape-DNA and Shape-DNA, respectively. To speed-up experiments, all shape signatures were computed offline, albeit their computation is quite inexpensive due in large part to the fact that only a relatively small number of eigenvalues of the LBO need to be calculated.

\begin{table}[!htb]
\caption{Classification accuracy results on the SHREC-2010 dataset. Boldface number indicates the best classification performance.}
\medskip
\centering
\begin{tabular}{lc}
\hline
\textbf{Method}    & \textbf{Average accuracy \%}\\
\hline
Shape-DNA  & 90.96 \\
cShape-DNA & 92.90 \\
GPS-embedding & 88.87 \\
F1-features & 86.49 \\
F2-features & 84.11 \\
F3-features & 87.72 \\
GA-BoF & 86.02 \\
SGWC-BoF & \textbf{95.66}\\
\hline
\end{tabular}
\label{Tab:SHREC10class}
\end{table}

\subsection{SHREC-2011 Dataset}
SHREC-2011 is a dataset of 3D shapes consisting of 600 watertight mesh models, which are obtained from transforming 30 original models~\cite{Lian:SHREC11}. Each shape in the dataset has approximately $m=1502$ vertices.

\begin{figure*}[!htb]
\centering
\includegraphics[scale=.97]{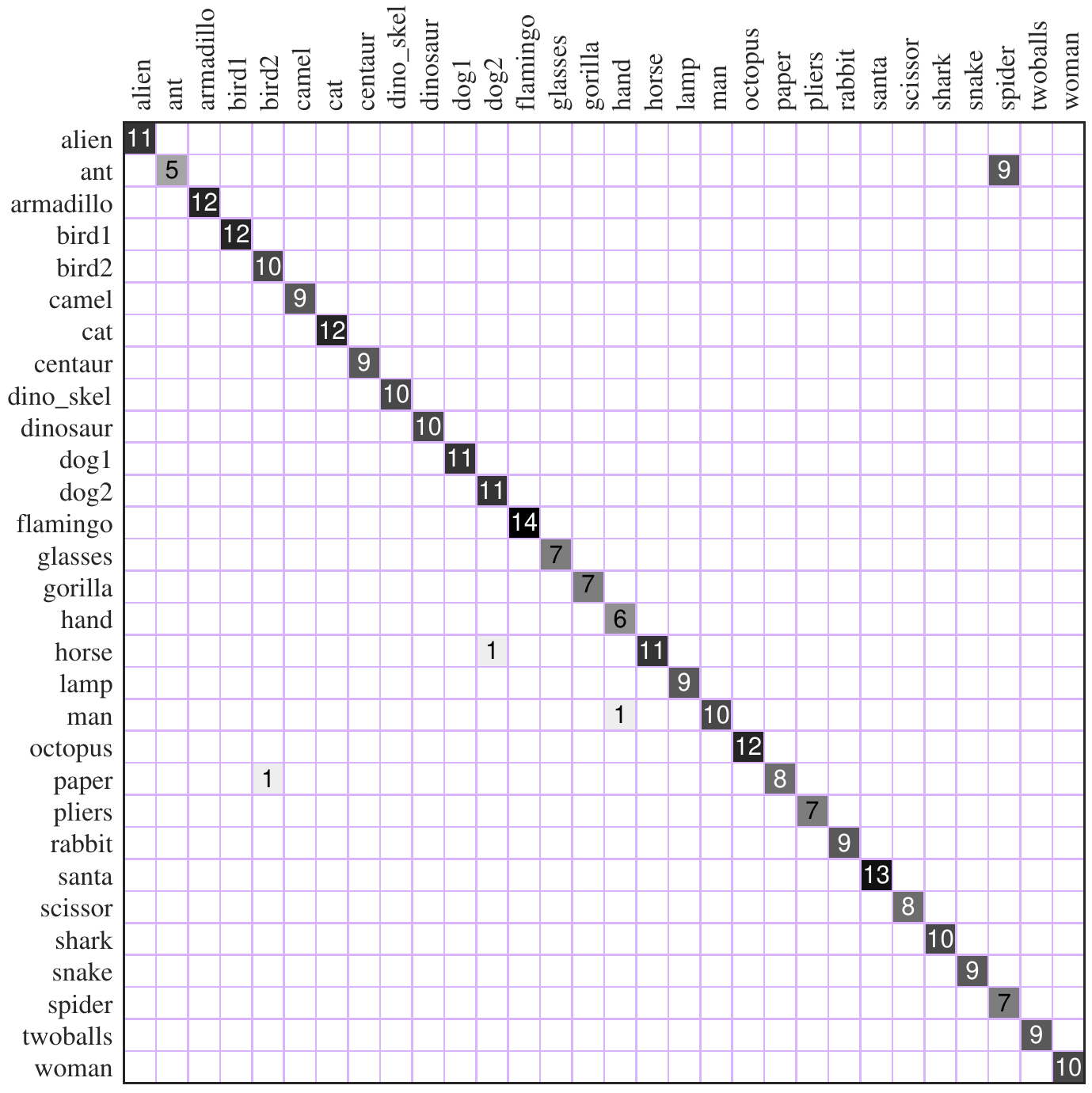}
\caption{Confusion matrix for SHREC-2011 using linear multiclass SVM.}
\label{Fig:confmatrix11}
\end{figure*}

\medskip
\noindent{\textbf{Performance Evaluation.}}\quad We randomly selected 50\% shapes in the SHREC-2011 dataset to hold out for the test set, and the remaining shapes for training. That is, the test data consists of 300 shapes. First, we trained a one-vs-all multiclass SVM on the training data to learn the classification model. Then, we used the resulting, trained model on the test data to predict the class labels. As can be seen in Figure~\ref{Fig:confmatrix11}, all shapes were classified correctly, except the horse, man and paper models, which were misclassified once as dog1, hand and bird1, respectively. Moreover, the ant shape was misclassified nine times as a spider. Therefore, the proposed  approach was able to accurately classify all shapes in the test data, as shown in Figure~\ref{Fig:confmatrix11}.

\medskip
\noindent{\textbf{Results.}}\quad Following the setting of the previous experiment, each 3D shape in the SHREC-2011 dataset is represented by a $5\times 1502$ spectral graph wavelet signature matrix. We pre-computed offline a vocabulary of size $k=128$, and it took about 70 minutes. The soft-assignment coding yields a $128\times 1502$ matrix $\bm{U}$ of mid-level features. Hence, the SGWC-BoF data matrix $\bm{X}$ for SHREC-2011 is of size $128^2\times 600$. We repeated the experimental process 10 times with different randomly selected training and test data in an effort to obtain reliable results, and the accuracy for each run was recorded. The average accuracy results are reported in Table~\ref{Tab:SHREC11class}. As can be seen, the proposed method performs the best compared to all the seven baseline methods. The highest classification accuracy of 97.66\% corresponds to our method, with performance improvements of 4.77\% and 3.25\% over the best performing baseline methods Shape-DNA and cShape-DNA, respectively.

\begin{table}[!htb]
\caption{Classification accuracy results on the SHREC-2011 dataset. Boldface number indicates the best classification performance.}
\medskip
\centering
\begin{tabular}{lc}
\hline
\textbf{Method}    & \textbf{Average accuracy \%}\\
\hline
Shape-DNA  & 92.89 \\
cShape-DNA & 94.41 \\
GPS-embedding & 88.40 \\
F1-features & 91.90 \\
F2-features & 89.47 \\
F3-features & 92.48 \\
GA-BoF & 93.20 \\
SGWC-BoF & \textbf{97.66}\\
\hline
\end{tabular}
\label{Tab:SHREC11class}
\end{table}

\subsection{Parameter Sensitivity}
\begin{figure*}[!htb]
\centering
\begin{tabular}{cc}
\includegraphics[scale=.6]{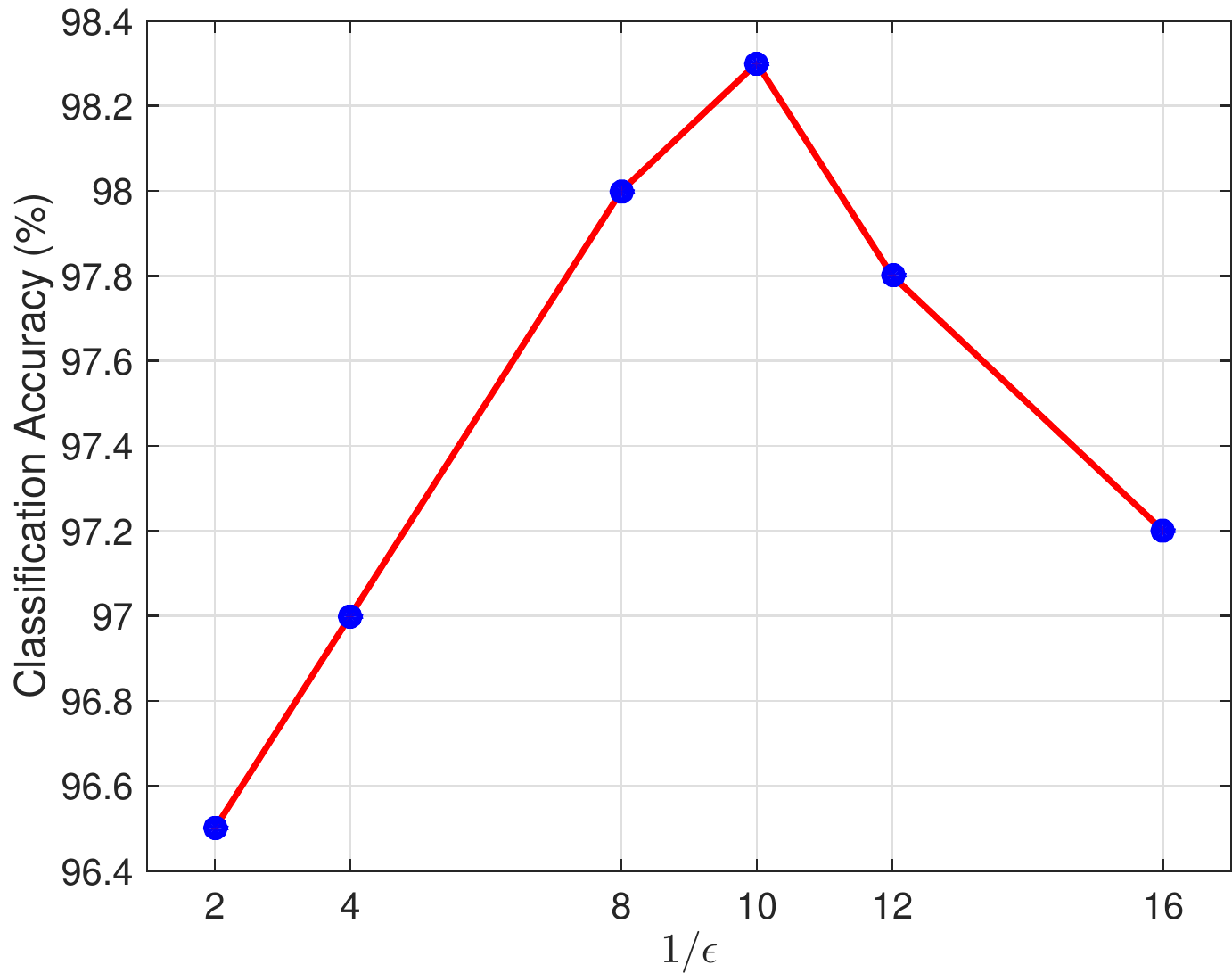} &
\includegraphics[scale=.6]{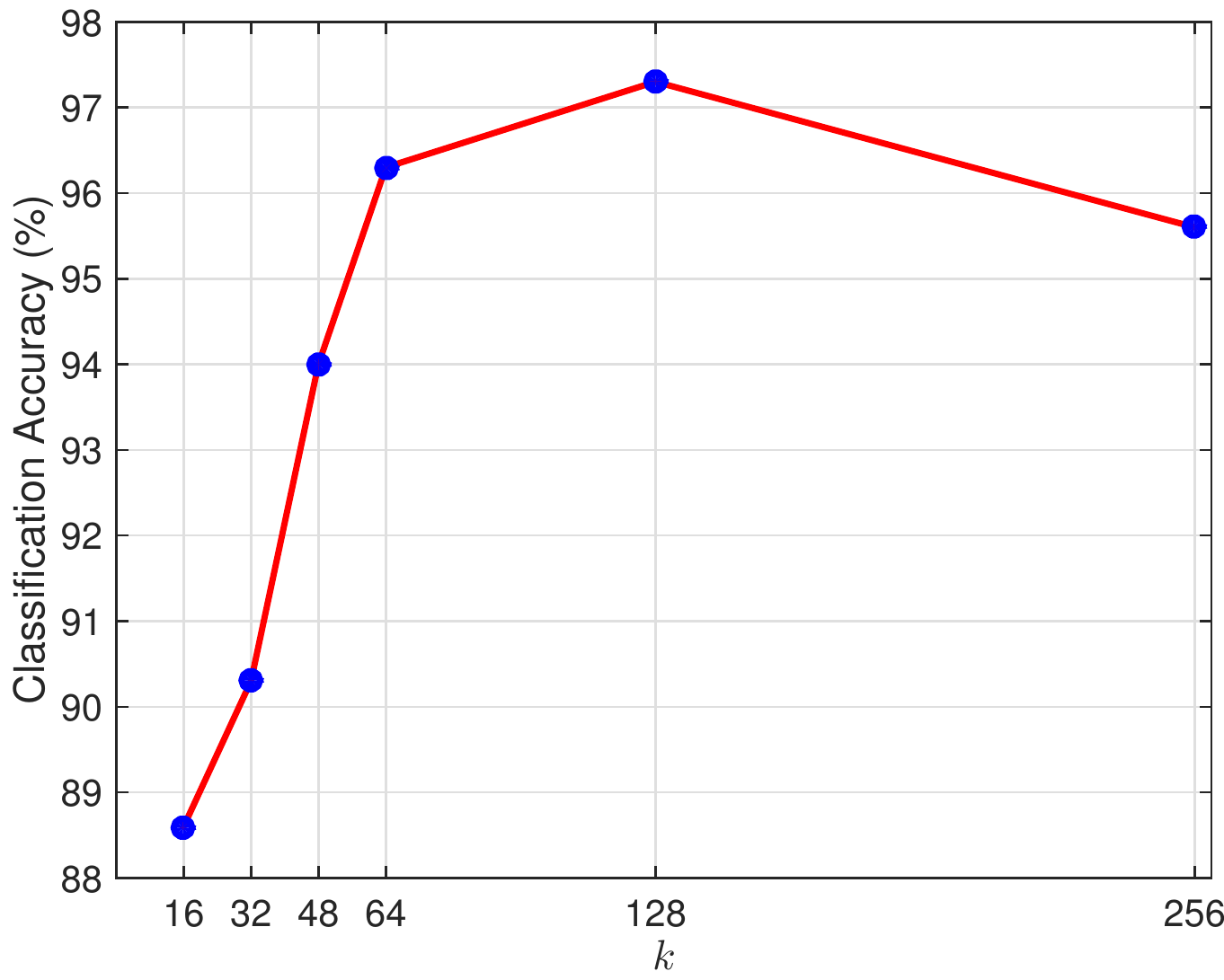}
\end{tabular}
\caption{Effects of the parameters on the classification accuracy for SHREC-2011.}
\label{Fig:parameters10}
\end{figure*}

The proposed approach depends on two key parameters that affect its overall performance. The first parameter is the kernel width $\epsilon$ of the geodesic exponential kernel. The second parameter $k$ is the size of the vocabulary, which plays an important role in the SGWC-BoF matrix $\bm{F}$. As shown in Figure~\ref{Fig:parameters10}, the best classification accuracy on SHREC-2011 is achieved using $\epsilon=0.1$ and $k=128$. In addition, the classification performance of the proposed method is satisfactory for a wide range of parameter values, indicating the robustness of the proposed framework to the choice of these parameters.

\section{Conclusion}
In this paper, we introduced a spectral graph wavelet framework for 3D shape classification that employs the bag-of-features paradigm in an effort to design a global shape descriptor defined in terms of mid-level features and a geodesic exponential kernel.  An important facet of our approach is the ability to combine the advantages of wave and heat kernel signatures into a compact yet discriminative descriptor, while allowing a multiresolution representation of shapes. The proposed spectral shape descriptor also combines the advantages of both band-pass and low-pass filters. In addition to taking into consideration the spatial relations between features via a geodesic exponential kernel, the proposed approach performs classification on spectral graph wavelet codes, thereby seamlessly capturing the similarity between these midlevel features. We not only showed that our formulation allows us to take into account the spatial layout of features, but we also demonstrated that the proposed framework yields better classification accuracy results compared to state-of-the-art methods, while remaining computationally attractive. This better performance is largely attributed to the discriminative global descriptor constructed by aggregating mid-level features weighted by a geodesic exponential kernel. Extensive experiments were carried out on two standard 3D shape benchmarks to demonstrate the effectiveness of the proposed method and its robustness to the choice of parameters. We evaluated the results using several metrics, including the confusion matrix and average accuracy. For future work, we plan to apply the proposed approach to other 3D shape analysis problems.

\bibliographystyle{ieeetr}
\bibliography{biblio}

\begin{thebibliography}{10}

\bibitem{Rosenberg:97}
S.~Rosenberg, {\em The Laplacian on a Riemannian Manifold}.
\newblock Cambridge University Press, 1997.

\bibitem{Levy:06}
B.~L\'evy, ``Laplace-{B}eltrami eigenfunctions: Towards an algorithm that
  ``understands'' geometry,'' in {\em Proc. {IEEE} Int. Conf. Shape Modeling
  and Applications}, p.~13, 2006.

\bibitem{Reuter:06}
M.~Reuter, F.~Wolter, and N.~Peinecke, ``Laplace-{B}eltrami spectra as
  '{S}hape-{DNA}' of surfaces and solids,'' {\em Computer-Aided Design},
  vol.~38, no.~4, pp.~342--366, 2006.

\bibitem{Rustamov:07}
R.~Rustamov, ``Laplace-{B}eltrami eigenfunctions for deformation invariant
  shape representation,'' in {\em Proc. Symp. Geometry Processing},
  pp.~225--233, 2007.

\bibitem{Bronstein:11}
A.~Bronstein, M.~Bronstein, L.~Guibas, and M.~Ovsjanikov, ``Shape {G}oogle:
  Geometric words and expressions for invariant shape retrieval,'' {\em ACM
  Trans. Graphics}, vol.~30, no.~1, 2011.

\bibitem{Chunyuan:13b}
C.~Li and A.~{Ben Hamza}, ``A multiresolution descriptor for deformable 3{D}
  shape retrieval,'' {\em The Visual Computer}, vol.~29, pp.~513--524, 2013.

\bibitem{Biasotti:SHREC17}
S.~Biasotti, E.~M. Thompson, M.~Aono, A.~B. Hamza, B.~B. stos, S.~Dong, B.~Du,
  A.~Fehri, H.~Li, F.~A. Limberger, M.~{Masoumi}, M.~Rezaei, I.~Sipiran,
  L.~Sun, A.~Tatsuma, S.~V. Forero, R.~C. Wilson, Y.~Wu, J.~Zhang, T.~Zhao,
  F.~Fornasa, and A.~Giachetti, ``{SHREC}'17 track: Retrieval of surfaces with
  similar relief patterns,'' in {\em Proc. Eurographics Workshop on 3D Object
  Retrieval 2017}, pp.~1--10, 2017.

\bibitem{Rodola:SHREC17}
E.~Rodola, L.~Cosmo, O.Litany, M.~M. Bronstein, A.~M. Bronstein, N.~Audebert,
  A.~B. Hamza, A.~Boulch, U.~Castellani, M.~N. Do, A.-D. Duong, T.~Furuya,
  A.~Gasparetto, Y.~Hong, J.~Kim, B.~L. Saux, R.~Litman, M.~{Masoumi},
  G.~Minello, H.-D. Nguyen, V.-T. Nguyen, R.~Ohbuchi, V.-K. Pham, T.~V. Phan,
  M.~Rezaei, A.~Torsello, M.-T. Tran, Q.-T. Tran, B.~Truong, L.~Wan, and
  C.~Zou11, ``{SHREC}'17 track: Deformable shape retrieval with missing
  parts,'' in {\em Proc. Eurographics Workshop on 3D Object Retrieval 2017},
  pp.~1--9, 2017.

\bibitem{Tarmissi:09}
K.~Tarmissi and A.~{Ben Hamza}, ``Information-theoretic hashing of {3D} objects
  using spectral graph theory,'' {\em Expert Systems with Applications},
  vol.~36, no.~5, pp.~9409--9414, 2009.

\bibitem{Gao:14}
Z.~Gao, Z.~Yu, and X.~Pang, ``A compact shape descriptor for triangular surface
  meshes,'' {\em Computer-Aided Design}, vol.~53, pp.~62--69, 2014.

\bibitem{Sun:09}
J.~Sun, M.~Ovsjanikov, and L.~Guibas, ``A concise and provably informative
  multi-scale signature based on heat diffusion,'' {\em Computer Graphics
  Forum}, vol.~28, no.~5, pp.~1383--1392, 2009.

\bibitem{Gebal:09}
K.~G{\c e}bal, J.~A. B{\ae r}entzen, H.~Aan{\ae s}, and R.~Larsen, ``Shape
  analysis using the auto diffusion function,'' {\em Computer Graphics Forum},
  vol.~28, no.~5, pp.~1405--1513, 2009.

\bibitem{Aubry:11}
M.~Aubry, U.~Schlickewei, and D.~Cremers, ``The wave kernel signature: A
  quantum mechanical approach to shape analysis,'' in {\em Proc. Computational
  Methods for the Innovative Design of Electrical Devices}, pp.~1626--1633,
  2011.

\bibitem{Kokkinos:10}
M.~Bronstein and I.~Kokkinos, ``Scale-invariant heat kernel signatures for
  non-rigid shape recognition,'' in {\em Proc. Computer Vision and Pattern
  Recognition}, pp.~1704--1711, 2010.

\bibitem{Chaudhari:14}
A.~Chaudhari, R.~Leahy, B.~Wise, N.~Lane, R.~Badawi, and A.~Joshi, ``Global
  point signature for shape analysis of carpal bones,'' {\em Physics in
  Medicine and Biology}, vol.~59, pp.~961--973, 2014.

\bibitem{Lian:13}
Z.~Lian, A.~Godil, B.~Bustos, M.~Daoudi, J.~Hermans, S.~Kawamura, Y.~Kurita,
  G.~Lavou{\'e}, H.~V. Nguyen, R.~Ohbuchi, Y.~Ohkita, Y.~Ohishi, F.~Porikli,
  M.~Reuter, I.~Sipiran, D.~Smeets, P.~Suetens, H.~Tabia, and D.~Vandermeulen,
  ``A comparison of methods for non-rigid {3D} shape retrieval,'' {\em Pattern
  Recognition}, vol.~46, no.~1, pp.~449--461, 2013.

\bibitem{Chunyuan:14b}
C.~Li and A.~{Ben Hamza}, ``Spatially aggregating spectral descriptors for
  nonrigid {3D} shape retrieval: A comparative survey,'' {\em Multimedia
  Systems, Springer}, vol.~20, no.~3, pp.~253--281, 2014.

\bibitem{Coifman:06}
R.~Coifman and S.~Lafon, ``Diffusion maps,'' {\em Applied and Computational
  Harmonic Analysis}, vol.~21, no.~1, pp.~5--30, 2006.

\bibitem{Hammond:11}
D.~Hammond, P.~Vandergheynst, and R.~Gribonval, ``Wavelets on graphs via
  spectral graph theory,'' {\em Applied and Computational Harmonic Analysis},
  vol.~30, no.~2, pp.~129--150, 2011.

\bibitem{Chunyuan:13c}
C.~Li and A.~{Ben Hamza}, ``Intrinsic spatial pyramid matching for deformable
  3d shape retrieval,'' {\em International Journal of Multimedia Information
  Retrieval}, vol.~2, no.~4, pp.~261--271, 2013.

\bibitem{Masoumi:16}
M.~Masoumi, C.~Li, and A.~B. Hamza, ``A spectral graph wavelet approach for
  nonrigid 3{D} shape retrieval,'' {\em Pattern Recognition Letters}, vol.~83,
  pp.~339--348, 2016.

\bibitem{Bu:14}
S.~Bu, Z.~Liu, J.~Han, J.~Wu, and R.~Ji, ``Learning high-level feature by deep
  belief networks for 3-{D} model retrieval and recognition,'' {\em IEEE Trans.
  Multimedia}, vol.~24, no.~16, pp.~2154--2167, 2014.

\bibitem{Masoumi:17}
M.~Masoumi and A.~B. Hamza, ``Spectral shape classification: A deep learning
  approach,'' {\em Journal of Visual Communication and Image Representation},
  vol.~43, pp.~198--211, 2017.

\bibitem{Meyer:03}
M.~Meyer, M.~Desbrun, P.~Schr\"oder, and A.~Barr, ``Discrete
  differential-geometry operators for triangulated 2-manifolds,'' {\em
  Visualization and mathematics III}, vol.~3, no.~7, pp.~35--57, 2003.

\bibitem{Wardetzky:07}
M.~Wardetzky, S.~Mathur, F.~K\"{a}lberer, and E.~Grinspun, ``Discrete {L}aplace
  operators: no free lunch,'' in {\em Proc. Eurographics Symp. Geometry
  Processing}, pp.~33--37, 2007.

\bibitem{Khabou:07}
M.~Khabou, L.~Hermi, and M.~Rhouma, ``Shape recognition using eigenvalues of
  the dirichlet laplacian,'' {\em Pattern Recognition}, vol.~40, pp.~141--153,
  2007.

\bibitem{Lian:SHREC10}
Z.~Lian, A.~Godil, T.~Fabry, T.~Furuya, J.~Hermans, R.~Ohbuchi, C.~Shu,
  D.~Smeets, P.~Suetens, D.~Vandermeulen, and S.~Wuhrer, ``{SHREC}'10 track:
  Non-rigid {3D} shape retrieval,'' in {\em Proc. Eurographics/ACM SIGGRAPH
  Sympo. 3D Object Retrieval}, pp.~101--108, 2010.

\bibitem{Lian:SHREC11}
Z.~Lian, A.~Godil, B.~Bustos, M.~Daoudi, J.~Hermans, S.~Kawamura, Y.~Kurita,
  G.~Lavoue, H.~Nguyen, R.~Ohbuchi, Y.~Ohkita, Y.~Ohishi, , F.~Porikli,
  M.~Reuter, I.~Sipiran, D.~Smeets, P.~Suetens, H.~Tabia, and D.~Vandermeulen,
  ``{SHREC}'11 track: Shape retrieval on non-rigid {3D} watertight meshes,'' in
  {\em Proc. Eurographics/ACM SIGGRAPH Symp. 3D Object Retrieval}, pp.~79--88,
  2011.

\end{thebibliography}
\end{document}